\documentclass[pra,twocolumn,amsmath,amssymb,floatfix]{revtex4}

\usepackage{amsmath}
\usepackage{graphicx}
\usepackage{epsfig}
\usepackage{hyperref}
\hypersetup{colorlinks=true}

\usepackage{bm}
\newcommand{\be}{\begin{equation}}
\newcommand{\ee}{\end{equation}}

\newcommand{\beq}{\begin{eqnarray}}
\newcommand{\eeq}{\end{eqnarray}}

\def\eq#1{(\ref{#1})}
\def\H1{\widehat{H}_1}

\begin{document}

\title[]{Quantum theory of light scattering in a one-dimensional channel: \\ Interaction effect on photon statistics and entanglement entropy}

\author{Mikhail Pletyukhov$^1$ and Vladimir Gritsev$^{2}$ }

\address{$^1$Institute for Theory of Statistical Physics and JARA -- Fundamentals of Future Information Technology, 
RWTH Aachen, 52056 Aachen, Germany \\
$^2$ Institute for Theoretical Physics, Universiteit van Amsterdam, Science Park 904,
Postbus 94485, 1098 XH Amsterdam, The Netherlands
}
\begin{abstract}
We provide a complete and exact quantum description of coherent light scattering  in a one-dimensional multi-mode transmission line coupled to a two-level emitter. Using recently developed scattering approach we discuss transmission properties, power spectrum, the full counting statistics and the entanglement entropy of transmitted and reflected states of light. Our approach takes into account spatial parameters of an incident coherent pulse as well as  waiting and counting times of a detector.  We describe time evolution of the power spectrum as well as observe deviations from the Poissonian statistics for reflected and transmitted fields. In particular, the statistics of reflected photons can change from sub-Poissonian to super-Poissonian for increasing values of the detuning, while the statistics of transmitted photons is strictly super-Poissonian in all parametric regimes. We study the entanglement entropy of some spatial part of the scattered pulse and observe that it obeys the area laws and that it is bounded by the maximal entropy of the effective four-level system.
\end{abstract}

\maketitle

\section{Introduction}
\subsection{Overview of studies of scattering in one-dimensional channel with emitter}

With current advances in experimental nanooptics, the problem of light scattering in quasi-one dimensional waveguides becomes an important cornerstone for understanding physics behind the light-matter interaction in a confined geometry. A number of recent experimental studies have been devoted to photon scattering when a single emitter is coupled to a one-dimensional (1D) scattering channel \cite{Akimov}, \cite{Astafiev}, \cite{Dayan}, \cite{coherent1}, \cite{coherent2}, \cite{chal1}, \cite{chal2}, \cite{Eichler1}, \cite{Eichler2}. The focus of these studies is made on a possibility of making few-photon devices (transistors, mirrors, switchers, transducers, etc.) as building blocks for either all-photonic or hybrid quantum devices.  While a number of few-photon emitters based on single molecules, diamond color centers and quantum dots are available nowadays \cite{Yamamoto},\cite{1-photon-rev}, an understanding of the extreme quantum regime of a few-photon scattering in a 1D fiber or transmission line \cite{Claudon},\cite{Imamoglu} should be supplemented by microscopic studies of scattering of a coherent light (e.g., generated by a laser driving) off an emitter in a confined 1D geometry. This is the main motivation of the present work. In addition, it is worth mentioning that the model studied here can be derived as an effective model in a 3D scattering geometry when  scattering channels are restricted to the photonic states with the lowest angular momentum values ($s$-wave scattering).   

Theoretical studies of quantum models describing light propagation in 1D geometry have been pioneered in 1980's by Rupasov and Yudson \cite{R1}, \cite{R2}, \cite{RY1}, \cite{RY2}, \cite{Y1}, \cite{Y2}, \cite{Y-entropy}. They introduced and solved a broad class of the Bethe ansatz integrable one-dimensional models, and even managed to determine exactly time evolution of the certain initial states \cite{Y1}. In the next decade these studies have extended by the other authors \cite{Le1}, \cite{KoLe}, \cite{LLLS}, \cite{Le2}. The exactly solvable  class of models includes linearly dispersed photons interacting with a single qubit, a Dicke cluster, and distributed emitters. However, the integrability imposes a rather strict constraint -- it requires the absence of backscattering thus limiting this class to the chiral, or unidirectional, models. This constraint, however, is not restrictive if a scatterer is local: transforming left- and right-propagating states of photons to the basis of their even and odd combinations, one can observe that the odd modes decouple from the scatterer and thus a model with backscattering is mapped onto an effective chiral model for even modes.  In turn, to realize the physical chiral model with distributed emitters  it has been recently proposed \cite{RPG} to employ scattering of  edge states in topological photonic insulators. Experimentally a quantum nondemolition measurement of a single
unidirectionally propagating microwave photon has been achieved in Ref.~\cite{QNDChalmers} using a chain of transmons cascaded through circulators which suppress photon backscattering.

A revival of interest to problems of photonic transport in 1D geometries has been triggered in 2000's by the progress in quantum information science, which resulted in series of publications from various groups \cite{Kojima1}, \cite{Kojima2}, \cite{Kojima3}, \cite{Shen-Fan1}, \cite{Shen-Fan2}, \cite{Shen-Fan3}, \cite{Shen-Fan4}, \cite{Chang}, \cite{Nori1}, \cite{YR}, \cite{Chang2}, \cite{Busch}, \cite{Sorensen}, \cite{Nori2}, \cite{Baranger1}, \cite{Roy}, \cite{Shi-Fan-Sun}, \cite{Hafezi1}, \cite{Baranger2}, \cite{Hafezi2}, \cite{Liao1}, \cite{Liao2}, \cite{PGnjp}, \cite{Oehri}, \cite{sanchez}, \cite{werra}, \cite{martens}, \cite{auf1}, \cite{auf2}. In these works, a variety of different setups have been carefully analyzed, comprising three- and four-level emitters, the nonlinear photon dispersion, effects of driving and dissipation. In addition, the recent experimental achievements \cite{Wallraff-2} motivate a theoretical consideration of models containing both distributed emitters and the backscattering \cite{Baranger3}, \cite{LP}, \cite{auf3}, \cite{2qubit-plasmon}.

\begin{figure*}[ht]
    \includegraphics[width=\textwidth]{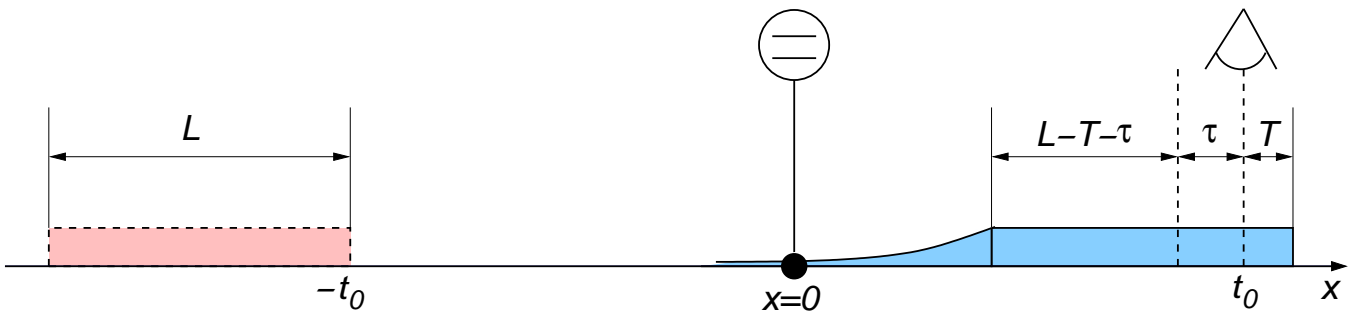}
    \caption{(Color online) Our system consists of a two-level emitter coupled to a waveguide (transmission line) at $x=0$. The sketch shows spatial snapshots of the wavepacket propagation. The coherent initial pulse $|\alpha_{0} \rangle$ of the length $L$ (shown in pink with dashed contour) is injected at time $t=-t_0$  and  the point $x=-t_0$. At time $t=0$ its front hits the scatterer. The scattered pulse (shown in blue with solid contour) leaves the scattering region, and after time $t_0$ its front reaches a detector located at $x=t_0$. At time $t=t_0 +T$ the detector starts counting photons which lasts during the time interval $\tau$. It is assumed that $t_0 \gg L > \tau$. } \label{fig:fcs}
\end{figure*}

\subsection{Scattering approach, role of detector}
In this paper we focus on a basic model consisting of a two-level qubit coupled to a 1D channel and driven by a coherent field. We develop a complete and exact quantum description of all physical properties of this system using the scattering formalism. 

A characterization of different scattering regimes in this model can be obtained by introducing parameters quantifying  (i) an initial state, (ii) a qubit (iii), a waveguide, and (iv) a detector. Throughout the paper we assume that the initial state is a pulse of the  spatial length $L$.  In the units $\hbar = v_g=1$, where $v_g$ is the group velocity of linearly dispersed photons in a waveguide, the parameter $1/L$ defines one of the important energy scales -- the wavepacket width in the $k$-space. Another energy scale is given by the qubit relaxation rate $\Gamma=\pi g^{2}$, where $g$ is a photon-qubit interaction strength. In addition, we have a dimensionless parameter $\bar{N}$ characterizing the mean number of photons in the initial pulse.  In terms of these parameters, one can distinguish three different regimes in this problem: (a) $\bar{N} \gg\Gamma L\gg 1$; (b) $\Gamma L\gg \bar{N} \gg 1$; (c) $\Gamma L\gg 1\gg \bar{N}$. Note that in all cases we assume $\Gamma L \gg 1$ meaning the long-$L$ (or narrow bandwidth) pulse's limit. 

The regime (a) was studied back in 1970's in connection with the resonance fluorescence phenomenon \cite{Mollow} (see also the review \cite{KM-rev}). In this regime, a semiclassical description of the laser beam is sufficient. 

In contrast to the regime (a), in the regime (c) single-photon (elastic) processes dominate, while  a contribution of many-photon (inelastic) processes to the scattering outcome is  weak; the most remarkable inelastic effect in this regime is, perhaps, a formation of the two-photon bound state. Various aspects of the regime (c) have been recently studied in numerous publications cited above. 

To complement the previous studies, we wish to achieve a comprehensive understanding of the crossover regime (b), where the mean number of scattered photons is already large, but the Rabi frequency $\propto \sqrt{\bar{N} \Gamma/L}$ is still much smaller than $\Gamma$.  To this end, we apply the quantum scattering approach developed in our earlier paper \cite{PGnjp}. It will be shown in the following that it is eventually capable to cover all three regimes, thereby establishing a theoretical platform for studying the classical-to-quantum crossover in this model.

In addition to the system  related parameters, our approach can accommodate information contained in the detection protocol as shown in Fig.~\ref{fig:fcs}. A pulse of the spatial length $L$ (shown in pink) is injected into the waveguide at time $t=-t_0$ and the coordinate $x=-t_0$ (recall that $v_g=1$). Due to the linear dispersion, it moves without changing its shape toward the qubit coupled to the waveguide at the point $x=0$. From the time instant $t=0$ the photons in the pulse start interacting with the qubit, and this interaction lasts approximately for a time $\sim L$. Subsequently the scattered pulse leaves the interaction region around $x=0$ (only the transmitted part is shown in blue in the figure), its shape is being modified, and at $t=t_0$ its front reaches the detector (the ``eye'') at the position $x=t_0$. To make the scattering formalism applicable we must assume that $t_0 \gg L$, meaning that the detector is far away from the scattering region. After the front of the scattered pulse reaches the detector at $t=t_0$, the latter remains switched off during the {\it waiting} time $T$, and at $t=t_0+T$ it switches on and starts to count transmitted photons during the subsequent time interval $\tau$ ({\it counting} time). Thus, the setup in Fig.~\ref{fig:fcs} is characterized by the two important time scales $T$ and $\tau$. In the following we will assume $L > \tau$ (which, in particular, allows us to take the limit $L \to \infty$ {\it before} the limit $\tau \to \infty$). This assumption is based on an instrumental possibility to create an initial wavepacket of a sufficiently narrow width in the $k$-space; from the conceptual viewpoint we impose this condition to enable an open-system description of the scattered pulse, for which the aforementioned order of limits is essential. 

The detector-related time scales together with the system-related energy scales exhaust the most relevant parameters of this problem.

\subsection{Physical observables and their generating function}
Major statistical properties of a scattered light field are  characterized by a collection of  $m$-point ($m=2,4, \ldots$) correlation functions \cite{Glauber}. 
Another basic observable quantity is a transmission (reflection) spectrum. It quantifies the amplitude of the transmitted (reflected) field, and it has been measured in various experiments with  microwave transmission lines \cite{Astafiev}, \cite{Dayan}, \cite{coherent1}, \cite{coherent2}. A more involved quantity of interest is the resonance fluorescence power spectrum, which is a Fourier transform of the field-field ($g^{(1)}$) correlation function. This (Mollow) spectrum is know to feature a three-peak structure \cite{Mollow}. Further experimental practice is to collect information about the density-density ($g^{(2)}$) correlation function using the Hanbury-Brown-Twiss setup \cite{HBT}. Higher order correlators can also (at least, in principle) be accessed in an experiment, and this motivates us to study the generating function of all  moments of the density operator, being related to the $m$-point functions. In mesoscopic physics, such a generating function is known as the  Full Counting Statistics (FCS), though in the 
historical retrospective this concept has been originally introduced in
the quantum-optical context \cite{Glauber}, \cite{MaWo} in order to characterize statistical
properties of a {\it non-interacting} quantized electromagnetic field. Its fully quantum derivation has
been presented in Ref.~\cite{KK}. For a coherent light field driving a two-level system (the resonance fluorescence model) the FCS has been studied in \cite{Mandel}, \cite{Cook}, \cite{Lenstra}, \cite{Smirnov-Troshin}. It has been observed \cite{Mandel} that the FCS distribution of the fluorescent photons is narrower than  the Poissonian distribution. An important measure of this effect is the Mandel's $Q$-parameter, $Q=  [\langle N^{(2)} \rangle  - \langle N^{(1)}\rangle^2 ]/\langle N^{(1)}\rangle$ which is obtained from the first and the second factorial moments of the FCS. 

Later on, the concept of the FCS has been borrowed and actively developed in
the field of mesoscopic physics \cite{LL}, \cite{LLL} for studying statistical properties of electronic currents
in meso- and nanoscopic devices for both non-interacting and interacting electrons
\cite{NB}, \cite{Belzig}, \cite{BN}, \cite{BUGS}, \cite{GK}, \cite{Schonh}, \cite{KS}, \cite{BSS}, \cite{BBSS}, \cite{CBS}. It has been recently
shown that the FCS can be very useful for characterizing classical-quantum crossover \cite{SB}, quantum
entanglement  \cite{TF}, \cite{KL}, \cite{LeHur}, and phase transitions \cite{IA}, \cite{FG}. The FCS
of nonlocal observables can be used to quantify correlations \cite{GADP}, \cite{IGD}, \cite{Hoffer}, \cite{LF}, and
prethermalization in many-body systems \cite{Kitagawa}, \cite{prethermalization}, as well as to define some kind of a topological order parameter \cite{IA2}. 

The studies of the FCS in mesoscopic physics have generated a back flow of ideas to the quantum optics community. Inspired by the recent experiments  in the context of the 1D resonance fluorescence, the subject of the FCS has received the renewed attention, see \cite{deflection}, \cite{BB}, \cite{Vogl}, \cite{LJ}.

\subsection{This paper: content and results}
Motivated by previous developments we revisit the original problem of computing the FCS for photons
interacting with an emitter. This is the first goal of the present work. We  give it a detailed quantum consideration, treating the interaction
nonperturbatively. We present an {\it exact} calculation of the FCS in the basic model of light-matter
interaction (see Fig.~\ref{fig:fcs}): a multimode propagating photonic field in 1D interacting with a
two-level emitter.  Since one of the objectives in nanophotonics research  is to
obtain strong photon nonlinearities as well as a strong photon-emitter interaction for the purpose
of an efficient control over individual atoms and photons, knowledge of
statistical properties of an {\it interacting} photon-emitter device becomes essential. 

The geometry of our system (see Fig.~\ref{fig:fcs}) suggests to define the three types of the counting statistics: the FCS of transmitted photons, the FCS of reflected photons, and the FCS of the chiral model. Below we discuss this classification in more detail. The multimode nature of the waveguide implies an emergence of many-body correlations. As such, interactions modify the statistics of photons in forward
and backward scattering channels in comparison with the Poissonian one of the incident coherent beam. As a by-product of our FCS computation, we revisit the transmission properties and evaluate the Mandel's $Q$-factor exhibiting super-Poissonian, sub-Poissonian, and Poissonian statistics for the three counting models in question.   

We also provide a new derivation of the Mollow spectrum based on knowledge of the exact scattering wavefunction that avoids the usage of the quantum regression theorem, and  reproduces the original expression \cite{Mollow} derived for the 3D scattering geometry. It helps to understand how the resonance fluorescence can be decomposed into elementary scattering processes.

Yet another quantity that has recently received a great deal of attention  due to developments in quantum information science is the {\it entanglement entropy}, see, e.g., the recent review \cite{Eisert}. While several measures of entanglement exist, the entropy of entanglement has
several nice properties like additivity and convexity. 
In quantum information theory, the entanglement entropy gives the efficiency of conversion of partially entangled to maximally entangled states by local operations \cite{Bennett1}, \cite{Bennett2}. In other terms, it gives the amount of classical information required to
specify the reduced density matrix. A large degree of entanglement is what makes quantum information exponentially more powerful than classical information, 
so states with lower entanglement entropy are less complex. For extended systems of condensed matter physics it is customary to distinguish between {\it area} and {\it volume} law \cite{Eisert} behavior of the entanglement entropy. Here the notions of area or volume refer to a typical geometric measure of a region bounded by a subsystem $A$ with respect to the rest of a system. Thus in 1D case, relevant for our discussion here, the area of an interval  $A$ consists of just two end points, while the volume is a length of the interval of the subsystem $A$. 
Systems with volume law behavior entanglement possess much higher potential for applications in quantum simulations and computing. It was shown \cite{Eisert} that in most cases a quantum ground state wave function of gapped systems exhibits the area law, while typical excited states mostly follow the volume law. An intermediate logarithmic behavior of the entanglement entropy is related to gapless systems. These features should be understood as asymptotic properties of a system, when the area and the volume of a subsystem entangled with the rest part of a system become large. Our complete knowledge of the scattering state allows us to calculate explicitly the entanglement entropy of the scattered pulse's interval of the length $\tau$ (see Fig.~\ref{fig:fcs}) for different values of $\tau$, $T$, and system parameters. The (dimensionless) duration $\Gamma\tau$ of the observation interval plays the role of the volume of the subsystem $A$ in this context.  One of our central results in that section is a demonstration of the existence of the  {\it absolute limit} for the entanglement entropy in our system: it is bounded by $\log 4$, the entropy of four-level system. Another important observation is that while the entanglement entropy at large $\tau$ asymptotically approaches the {\it area law} bounded by $\log 4$, it can behave very differently for small and intermediate values of $\tau$ --  we even observe its nonmonotonous oscillatory behavior for some intermediate regime of parameters.

\section{Definitions and approximations}
We start our analysis defining the model and approximations involved in its derivation, the bosonic operators creating the initial pulse, and the FCS. 

\subsection{Theoretical model.}

\subsubsection{Approximations and the effective Hamiltonian.}

Our model is described by the Hamiltonian $H=H_{ph}+H_{em}+H_{ph-em}$, where $H_{ph}$ is the Hamiltonian of the free propagating photonic field,  $H_{em}$ is the Hamiltonian of an emitter, and $H_{ph-em}$ describes the field-emitter coupling. We involve approximations which are customary in quantum optics: (i) the dipole approximation for the interacting Hamiltonian; (ii) the two-level approximation for the emitter Hamiltonian; (iii) the rotating wave approximation (RWA); and (iv) the Born-Markov approximation (energy independence) for the coupling constant. In addition, we linearize the photonic spectrum around some appropriately chosen frequency $\Omega_{0}$ which is commensurate with the emitter's transition frequency $\Omega$, and extend the linearized dispersion to infinities. With these assumptions [except for (iii)] we obtain an effective low-energy Hamiltonian
\beq
H&=&\sum_{\xi=r,l}\int dk(\Omega_{0}+\xi k)a^{\dag}_{\xi k}a_{\xi k}+\frac{\Omega}{2}\sigma^{z}\nonumber\\
&+&g_0 \sum_{\xi=r,l}\int dk (a_{\xi k}^{\dag}+a_{\xi k})(\sigma^{+}+\sigma^{-}),
\eeq
featuring the two-branch linear dispersion with right- ($\xi=r=+$) and left- ($\xi=l=-$) propagating modes. Here 
$\sigma^{\pm}=(\sigma^{x}\pm i\sigma^{y})/2$  are expressed in terms of the Pauli matrices $\sigma^{a}$, $a=x,y,z$. The states of the emitter are separated by the transition frequency $\Omega$. To implement the RWA in a systematic way, we first perform the gauge transformation $H \to U^{\dag} H U  + i (d U^{\dag}/ dt) U$ with 
\beq
U=\exp\left[-i\Omega_{0}t\left(\sum_{\xi=r,l}\int dk \, a^{\dag}_{\xi k} a_{\xi k}+\frac{\sigma^z}{2} \right)\right],
\eeq
which leads us to the Hamiltonian
\beq
H&=&\sum_{\xi=r,l}\int dk \, \xi \, k \, a^{\dag}_{\xi k}a_{\xi k}+\frac{\Delta}{2}\sigma^{z}\nonumber\\
&+&g_0 \sum_{\xi=r,l}\int dk (a_{\xi k}^{\dag}\sigma^{-}+a_{\xi k}\sigma^{+}) \label{start-Ham} \nonumber\\
&+& g_0 \sum_{\xi=r,l}\int dk (a_{\xi k}^{\dag}\sigma^{+} e^{2i\Omega_{0}t}+a_{\xi k}\sigma^{-} e^{-2i\Omega_{0}t}),
\label{osc-terms}
\eeq
where $\Delta=\Omega-\Omega_{0}$. As soon as $g_0^2/\Omega_0 \ll 1$, the time-oscillating terms in (\ref{osc-terms}) can be treated as a time-dependent perturbation. In zeroth order it is simply neglected, which is equivalent to the RWA. Note that this approximation is consistent with the assumption about the absence of lower and upper bounds in the linearized dispersion.

\subsubsection{Transformation to the ``even-odd'' basis.}
Due to energy independence of the coupling constant $g_0$, one can decouple the Hilbert space of the model defined in (\ref{start-Ham}) into two sectors. To this end, one introduces \emph{even} (symmetric) and \emph{odd} (antisymmetric) combinations of fields corresponding to the same energy $|k|$
\be
a_{ek} = \frac{a_{rk} + a_{l, -k}}{\sqrt{2}}, \quad a_{ok} = \frac{a_{rk}-  a_{l, -k}}{\sqrt{2}} .
\ee
By virtue of this canonical transformation the Hamiltonian (\ref{start-Ham}) turns into a sum of the two terms, $H=H_e + H_o$, defined by
\beq
H_e &=& \int dk \left[ k \, a^\dag_{e k} a_{e k} 
+ g  \left( a^\dag_{e k}\sigma^-  + a_{e k}\sigma^+ \right) \right]
+ \frac{\Delta}{2} \sigma^z , \nonumber\\
H_o &=&  \int dk \, k \,  a^\dag_{ok} a_{ok},
\eeq
where $g=g_0 \sqrt{2}$. Note that the odd Hamiltonian $H_{o}$ is noninteracting, and therefore  odd modes do not scatter off a local emitter ($S_o \equiv 1$). The even Hamiltonian $H_{e}$ can be interpreted in terms of a chiral model with a single branch of the linear dispersion. 

A similar decomposition can be applied to an initial state. Both even and odd photons are labeled by a momentum value $k$ lying on a single branch of the linear dispersion.

\subsection{Definitions of wave packet field operators and the initial state.}

In order to define the incident coherent state we need field operators annihilating/creating states which are normalized by unity. The field operators $a_{k}$ and $a^{\dag}_{k}$, which are the Fourier transforms of $a (x)$ and $a^{\dag} (x)$, do not fulfill this requirement, as they obey the commutation relation $[a_k, a^{\dagger}_{k'}] = \delta (k-k')$ (in other words, they annihilate and create {\it unnormalizable} states). To circumvent this difficulty, we construct wavepacket field operators
\beq
b_k^{\dagger} &=& \frac{1}{\sqrt{L}} \int_{-L/2}^{L/2} d x a^{\dagger} (x ) e^{i k x},
\label{op_c}
\eeq
which do satisfy the desired commutation relation $[b_{k}, b_{k}^{\dagger}] =1$. For example, the operators 
$b_{r,k_0}^{\dagger}   =  \frac{1}{\sqrt{L}} \int_{-L/2}^{L/2} d x a^{\dagger}_{r} (x) e^{i k_0 x} $ and $b_{l,-k_0}^{\dagger}  = \frac{1}{\sqrt{L}} \int_{-L/2}^{L/2} d x a^{\dagger}_{l} (x) e^{-i k_0 x}$ create wavepackets which are centered around $+k_0 (-k_0)$ of the right (left) branch of the spectrum and broadened over the width $\sim 2 \pi/L$. In the coordinate representation, they create states which are spatially localized on a finite interval of the length $L$. We note the identity $[a_{\xi} (x), b_{\xi',k_0}^{\dagger}] = \delta_{\xi \xi'} \frac{e^{i k_0 x}}{\sqrt{L}} \Theta (L/2-|x|)$. In the following, $k_0$ denotes the laser driving frequency (measured from the linearization point).

Having introduced $b_{r/l,k0}^{\dagger}$, we define the initial (incoming)
state $|\mathrm{in} \rangle =|\alpha_0 \rangle_{r} \otimes |\downarrow \rangle$, where the
incident right-moving photons are prepared in the coherent state $|\alpha_0 \rangle_{r}  =  D_{r} (\alpha_0) |0 \rangle$, and the two-level emitter is initially in the
ground state~$|\downarrow\rangle$. Here $|0\rangle$~denotes the photonic vacuum, and $D_{r} (\alpha_0) = \exp (\alpha_0 b^{\dagger}_{r, k_0} - \alpha_0^* b_{r, k_0}) $ is the coherent state displacement operator. 

The mean number of photons
in the state $|\alpha_0 \rangle_{r}$ is given by $\bar{N}_0 = |\alpha_0 |^2$. We also quote a useful relation
\beq
  D_{r}^{\dagger} (\alpha_0) a_r (x ) D_{r} (\alpha_0) = a_r (x) + \frac{\alpha_0 e^{i k_0 x}}{\sqrt{L}} \Theta (L/2 - |x|),
 \label{til_a}
\eeq
which follows from the commutation relation between $a$  and $b^{\dagger}$ operators.

The initial coherent state defined in the original right-left basis admits a decomposition into the product state in the even-odd basis
\beq
|\alpha_0 \rangle_{r}&=&e^{\alpha_0 b^{\dag}_{r, k_0} -|\alpha_0|^2/2} | 0 \rangle \nonumber\\
&=& 
e^{(\alpha_0/\sqrt{2}) b^{\dag}_{e,k_0} -|\alpha_0|^2/4} e^{(\alpha_0/\sqrt{2}) b^{\dag}_{o,k_0} - |\alpha_0|^2/4} | 0 \rangle \nonumber\\
&\equiv& D_e (\alpha )D_{o} (\alpha ) |0\rangle\equiv |  \alpha \rangle_e \otimes | \alpha \rangle_o ,\nonumber
\label{init-state}
\eeq
where $\alpha = \alpha_0/\sqrt{2}$, and the displacement operators $D_{e,o}$ are defined using the mutually commuting operators $b_{e, k_{0}}^{\dagger}$ and $b_{o,k_{0}}^{\dagger}$, respectively. Importantly, $\, _r\langle \alpha_0 | \alpha_0 \rangle_r =1$, i.e. the incoming state is properly normalized.

The major consequence of the even-odd decoupling is a factorization of the scattering operator $S$
into the product $S= S_e S_o$, where $S_o$ is the identity operator, and $S_e$ can be studied in the context of the effective one-channel chiral model described by $H_e$. Scattering in chiral models has been studied by us in Ref.~\cite{PGnjp} for arbitrary initial states, including the coherent state. In particular, we have established the explicit form of the operator $S_e$ in the latter case, which provides the full information about the scattering wavefunction. Here we take over this result and use  it for calculation of observables announced in the Introduction. All expressions necessary for this purpose are quoted below for readers' convenience.

\subsection{Definition of the full counting statistics}

The statistics of the initial field, defined by the probability $p_{\alpha_0} (n)$ to find $n$ photons in
the mode $k_0$, is given by the Poissonian distribution $p_{\alpha_0} (n)= e^{-\bar{N}_0}
\frac{\bar{N}_0^{n}}{n!}$ for the coherent field, with the mean value $\bar{N}_0 =|\alpha_0|^2$.  Due to the photonic dispersion  and inelastic scattering processes photons can leak from the right-moving
mode $k_0$ to other modes on both branches of the spectrum by virtue of scattering processes, thus modifying the photon statistics. 
A fraction of photons is reflected, and their statistics is also of great
interest. We propose a calculation of the FCS in both forward and backward scattering channels,
which is {\it exact} and {\it nonperturbative} in both $g_0$ and $\bar{N}_0$.

Generally speaking, the FCS can be defined as a  generating function $F(\chi)
=\sum_{n=0}^{\infty} e^{i \chi n} p (n)$ associated with  a probability  distribution $p(n)$ to detect $n$ photons in some given state. The function $F (\chi) $
generates $m$-th order moments of the distribution $p (n)$ which are determined by evaluating the $m$-th derivative with respect to $i \chi$ at $\chi=0$. 
The Fourier expansion of the $2\pi$-periodic function $F(\chi)$ yields power series in terms of
the ``fugacity'' $z=e^{i\chi}$, $F(\chi) = \sum_{n=0}^{\infty} z^{n} p(n) \equiv {\hat
F}(z)$. The normalization of a probability distribution implies $\hat F(1) = 1$. The expansion  $\hat{F} (z)= \sum_{r=0}^{\infty} \frac{(z-1)^r}{r!} \langle N^{(r)} \rangle$ around $z=1$ gives the factorial moments of a distribution $\langle N^{(r)} \rangle = \langle N (N-1) \ldots (N-r+1)\rangle$.

To define the photon FCS  we need the two main constituents: (i) the scattering  (outgoing) state $|\mathrm{out}\rangle = (S_e | \alpha \rangle_e ) \otimes |\alpha \rangle_o $ necessary to perform the average,  and (ii) a meaningful and experimentally measurable counting operator $N$. In particular, as such we can choose the number of transmitted photons which pass through the detector during the time $\tau$ (see the Fig.~\ref{fig:fcs}). Because of the linear dispersion, the same operator characterizes the number of photons in the spatial interval of the length $v_g \tau$ viewed in the frame co-moving in the right direction with the velocity $v_g$. Introducing the coordinate system in this co-moving frame such that the pulse's front has the coordinate value $+L/2$, we define the photon number operator
\beq
N_{r, \tau} = \int_{z_1}^{z_2} d x a_r^{\dagger} (x) a_r (x) 
\label{Ndef}
\eeq
of transferred photons which appear in the spatial interval $[z_1,z_2]$. Here $-L/2<z_1<z_2 <L/2$ (we note that the tail of the scattered pulse extends to $-\infty$, see below; however, we will focus on counting intervals $\tau \equiv z_2 -z_1 < L$). The corresponding FCS reads
\beq
F_{r, \tau} (\chi_0) = \langle e^{i \chi_0 N_{r,\tau}} \rangle .
\label{Fdef}
\eeq
In the chosen coordinate system, the waiting time is expressed by $T =L/2 -z_2$. 

In the same setting one can consider the FCS of the chiral model.

To define the FCS for reflected photons, one can put the second detector at $x=-t_0$ and consider the co-moving frame with the velocity $-v_g$, defining in it the counting operator $N_{l,\tau}$ via $a_l (x)$.

\section{Method of computing observables based on exact scattering matrix}

\subsection{Scattering of the coherent state in the chiral model}

As we discussed above, we need to know an expression $S_e | \alpha \rangle_e$ for the scattering state in the effective chiral model for the even modes. In this subsection we quote the result of Section 5 in Ref.~\cite{PGnjp} for the coherent light scattering in the chiral model. In particular, we  copy the Eq.~(134) from this reference, adapting notations therein to the present paper.
The subscript $e$ is also omitted in the following.

Thus, for the incoming coherent state $| \alpha \rangle$ in the mode $k_0$, the outgoing scattering state amounts to $S_0 | \alpha \rangle$, where
\beq
S_0 = S_0^a [L/2,-L/2] + S_0^b [L/2,-L/2] \frac{\lambda}{\sqrt{2 \Gamma}} A_0^{\dagger}.
\label{S0-factor}
\eeq

The operator
\be
A_0^{\dagger} =  \sqrt{2 \Gamma} \int_{-\infty}^{-L/2} d x_0 e^{i k_0 x_0} e^{- i (\delta + i \Gamma) (L/2 +x_0)} a^{\dagger} (x_0),
\ee
describing the states in the tail of the scattered pulse, is normalized by $\langle 0 | A_0 A_0^{\dagger}  | 0 \rangle =1$. 

In turn, the states within the initial pulse's size $L$ are expressed via the operators 
\beq
S_0^a [y,x]  &=& 1 + \sum_{n=1}^{\infty} \lambda^n \int {\cal D}x_{n}\label{Sa0}\\
& \times & d_0 (y - x_n) a^{\dagger} (x_n) e^{i k_0 x_n}\nonumber\\
&\times& d_0 (x_n - x_{n-1}) a^{\dagger} (x_{n-1})e^{i k_0 x_{n-1}} \nonumber\\
&\times & \ldots d_0 (x_2 -x_1) a^{\dagger} (x_1 ) e^{i k_0 x_1} ,  \nonumber\\  
S_0^b [y,x] &=& d_0 (y-x) + \sum_{n=1}^{\infty} \lambda^n \int {\cal D}x_{n}\label{Sb0}\\
& \times & d_0 (y - x_n) a^{\dagger} (x_n) e^{i k_0 x_n}\nonumber\\
&\times & d_0 (x_n - x_{n-1})
a^{\dagger} (x_{n-1}) e^{i k_0 x_{n-1}} \nonumber \\
&\times&  \ldots d_0 (x_2 - x_1) a^{\dagger} (x_1 ) e^{i k_0 x_1} d_0 (x_1 -x). \nonumber
\eeq
Here the parameter
\beq
\lambda = - \frac{2 i \Gamma \alpha}{(\delta + i \Gamma) \sqrt{L}},
\eeq
is expressed in terms of the detuning $\delta = k_0 - \Delta$ and the relaxation rate $\Gamma = \pi g^2$; $d_0 (x)=1-\exp [i (\delta + i \Gamma) x]$ is the bare single-photon propagator, and the short-hand notation has been used for the integration measure ${\cal D}x_{n} = \Theta (y > x_n > \ldots > x_1 >  x) d x_n \ldots d x_1 $. We also consider $L/2 \geq y >x \geq -L/2$.

For the later use we also define the following operators
\beq
S_0^c [y,x]  &=& 1+ \sum_{n=1}^{\infty} \lambda^n \int {\cal D}x_{n}\label{Sc0}\\
& \times &  a^{\dagger} (x_n) e^{i k_0 x_n} d_0 (x_n - x_{n-1}) \nonumber\\ 
&\times& a^{\dagger} (x_{n-1})e^{i k_0 x_{n-1}} d_0 (x_{n-1} -x_{n-2})\ldots \nonumber\\
&\times& d_0 (x_2 -x_1) a^{\dagger} (x_1 ) e^{i k_0 x_1} ,  \nonumber
\eeq
\beq
S_0^{\bar{a}} [y,x] &=& 1 + \sum_{n=1}^{\infty} \lambda^n \int {\cal D}x_{n} \label{Sat0} \\
& \times &  a^{\dagger} (x_n) e^{i k_0 x_n} d_0 (x_n - x_{n-1})\nonumber\\
&\times& a^{\dagger} (x_{n-1}) e^{i k_0 x_{n-1}}d_0 (x_{n-1} - x_{n-2}) \ldots \nonumber \\
&\times& a^{\dagger} (x_1 ) e^{i k_0 x_1} d_0 (x_1 -x).
\nonumber
\eeq
 The set of operators $S^{a}_{0},S^{b}_{0},S^{c}_{0},S^{\bar{a}}_{0}$ is complete in that sense that they exhaust all possible arrangements of the bare propagators $d_{0}(x)$. 

\subsection{Algebra of scattering operators}

The scattering state (134) of Ref.~\cite{PGnjp} (equivalent of $S_0 | \alpha \rangle$) can be used for a computation of observable quantities. In particular, we will be interested in correlation functions $\langle \alpha |S_0^{\dagger} a^{\dagger} (z_1) \ldots a^{\dagger} (z_m) a(z_m) \ldots a(z_1) S_0 | \alpha \rangle$, and therefore we need to know how the local annihilation operators $a (z_k)$ commute with the many-body scattering operator $S_0$ defined on a finite spatial interval. For a systematic treatment, we observe the following algebraic properties of the scattering operators $S^{a}_{0},S^{b}_{0},S^{c}_{0},S^{\bar{a}}_{0}$.

Let us choose an arbitrary point $z \in [x, y]$. Using the obvious identity
\beq
& &\Theta (y > x_n > \ldots > x_1 > x) \\
&=& \sum_{j=0}^n \Theta (y > x_n > \!\ldots >\! x_{j+1} > z > x_j > \!\ldots > \!x_1 >  x),\nonumber
\eeq
where $x_{n+1}=y$ and $x_0=x$, one can show by rewriting  Eqs.~\eq{Sa0}, \eq{Sb0}, \eq{Sc0}, and \eq{Sat0}, that the operators $S^{a}_{0},S^{b}_{0},S^{c}_{0},S^{\bar{a}}_{0}$ satisfy the following  closed algebra with respect to the interval splitting operation
\beq
S_0^a [y,x] \!=\! S_0^a [y,z] S_0^a [z,x]\! + S_0^b [y,z] \! \{ S_0^c [z,x]\!- \!S_0^a [z,x] \},\nonumber\\
S_0^b [y,x] \!=\! S_0^a [y,z] S_0^b [z,x]\! + S_0^b [y,z] \! \{ S_0^{\bar{a}} [z,x]\!  -\! S_0^b [z,x] \}, \nonumber\\
S_0^c [y,x] \!= \! S_0^c [y,z] S_0^a [z,x]\! + S_0^{\bar{a}} [y,z] \! \{S_0^c [z,x] \!-\!S_0^a [z,x]\}, \nonumber\\
S_0^{\bar{a}} [y,x]\! =\!   S_0^c [y,z] S_0^b [z,x] \! + S_0^{\bar{a}} [y,z] \! \{ S^{\bar{a}} [z,x]\! - \! S_0^b [z,x]\}.\nonumber\\
\label{split_alg}
\eeq
If one divides the interval $[x,y]$ into three parts $y>z_2>z_1>x$ by arbitrary points $z_{1}$ and $z_{2}$, one can prove by a direct calculation that the algebra \eq{split_alg} is associative, as expected. 

The algebra \eq{split_alg} also allows us to express the action of annihilation operators on the scattered state in a simple form
\beq
& &a (z) S_0^a [y,x] |0 \rangle = S_0^b [y,z] \lambda e^{i k_0 z} S_0^a [z,x] |0 \rangle,\label{aS0a}\\
& &a (z) S_0^b [y,x] | 0 \rangle = S_0^b [y,z] \lambda e^{i k_0 z} S_0^b [z,x] |0 \rangle, \label{aS0b}\\
& &a (z) S_0^c [y,x] |0 \rangle=  S_0^{\bar{a}} [y,z] \lambda e^{i k_0 z} S_0^a [z,x] |0 \rangle, \label{aS0c} \\
& &a (z) S_0^{\bar{a}} [y,x] |0 \rangle =  S_0^{\bar{a}} [y,z]  \lambda e^{i k_0 z} S_0^{b} [z,x] | 0 \rangle .
\label{aS0abar}
\eeq
For the proof of \eq{aS0a}-\eq{aS0abar} we used the property $d_0 (0) =0$.

Similarly, for an action of the ordered product of two annihilation operators $a(z_{2})a(z_{1})$,  with $y>z_{2}>z_{1}>x$, we obtain
\beq
& & a (z_2) a (z_1) S_0^a [y,x] |0 \rangle \label{aaS0a} \\
&=& S_0^b [y, z_2] \lambda e^{i k_0 z_2} S_0^b [z_2, z_1] \lambda e^{i k_0 z_1} S_0^a [z_1, x] |0 \rangle ,
\nonumber \\
& & a (z_2 ) a (z_1) S_0^b [y,x] |0 \rangle \label{aaS0b} \\
&=& S_0^b [y, z_2] \lambda e^{i k_0 z_2}S_0^b [z_2, z_1] \lambda e^{i k_0 z_1} S_0^b [z_1, x] |0 \rangle,
\nonumber  \\
& & a (z_2) a (z_1) S_0^c [y,x] |0 \rangle  \label{aaS0c} \\
&=& S_0^{\bar{a}} [y, z_2] \lambda e^{i k_0 z_2} S_0^b [z_2 , z_1] \lambda e^{i k_0 z_1} S_0^a [z_1, x] |0 \rangle, 
\nonumber  \\
& & a (z_2) a (z_1) S_0^{\bar{a}} [y,x]  |0 \rangle \label{aaS0abar} \\
&=&  S_0^{\bar{a}} [y, z_2] \lambda e^{i k_0 z_2} S_0^b [z_2,z_1] \lambda e^{i k_0 z_1} S_0^b [z_1, x] |0 \rangle. \nonumber
\eeq

Iterating this procedure, one can find an action of the ordered product of $m$ annihilation operators $a (z_m) \ldots a (z_1)$, $y>z_m>\ldots z_1>x$. It produces the product of $m+1$ $S$-operators, what can be symbolically written as
\beq
a^m S_0^a & \to & S_0^b \ldots S_0^b \ldots S_0^a , \label{ama} \\
a^m S_0^b & \to & S_0^b \ldots S_0^b \ldots S_0^b , \label{amb} \\
a^m S_0^c & \to & S_0^{\bar{a}} \ldots S_0^b \ldots S_0^a , \label{amc} \\
a^m S_0^{\bar{a}} & \to & S_0^{\bar{a}} \ldots S_0^b \ldots S_0^b . \label{ambara}
\eeq
Note that in all intermediate positions appears only $S_0^b$. To classify leftmost and rightmost operators in these expressions, we introduce the mappings $\sigma$ and $\mu$ according to the table
\beq
\sigma (a) =& \! b , \quad  \mu (a) =& \!a , \\
\sigma (b) =& \! b , \quad \mu (b) =& \! b , \\
\sigma (c) =& \! \bar{a}, \quad \mu (c) =& \! a , \\
\sigma (\bar{a}) =& \! \bar{a} , \quad \mu (\bar{a}) =& \! b .
\eeq
In their terms, the relations \eq{ama}-\eq{ambara} acquire the compact form
\beq
a^m S_0^{\beta} \to S_0^{\sigma (\beta)} \left( S_0^b \right)^{m-1} S_0^{\mu (\beta)}.
\label{comp_am}
\eeq 

\subsection{Dressing $S$-operators}
In the following we will also need {\it shifted} scattering operators
\beq
S_v^{\beta} [y, x] &=&  D^{\dagger} (v) S_0^{\beta} [y,x] D (v) \nonumber\\
&=& e^{v^* b_{k_0}} S^{\beta} [y,x]  e^{-v^* b_{k_0}} ,
\label{disp}
\eeq
where $\beta=a, b, c,\bar{a}$, and $D(v)= \exp [v b_{k_0}^{\dagger} - v^* b_{k_0}]$ is the displacement operator of the fields,  such that $D^{\dagger} (v) a^{\dagger} (x) D (v) = a^{\dagger} (x) + v^* e^{-i k_0 x}/\sqrt{L}$. Our next goal is to establish explicit expressions for the operators $S_v^{\beta} [y, x]$ for arbitrary complex-valued parameter $v$.

Performing the displacement \eq{disp}, we obtain the new series in field operators defining $S_v^{a, b,c,\bar{a}}$. Appropriately reorganizing (re-summing) them, we find the following expressions 
\beq
S_v^a [y,x]  &=& \tilde{d}_v (y-x) + \sum_{n=1}^{\infty} \lambda^n \int {\cal D}x_{n} \label{Sa}  \\
& \times & d_v (y - x_n) a^{\dagger} (x_n) e^{i k_0 x_n} \nonumber\\
&\times& d_v (x_n - x_{n-1}) a^{\dagger} (x_{n-1})e^{i k_0 x_{n-1}}  \nonumber\\
&\times & \ldots d_v (x_2 -x_1) a^{\dagger} (x_1 ) e^{i k_0 x_1} \tilde{d}_v (x_1 -x), 
\nonumber
\eeq
\beq
S_v^b [y,x] &=& d_v (y-x) + \sum_{n=1}^{\infty} \lambda^n \int {\cal D} x_{n} \label{Sb}
\\
& \times & d_v (y - x_n) a^{\dagger} (x_n) e^{i k_0 x_n} \nonumber\\
&\times& d_v (x_n - x_{n-1}) a^{\dagger} (x_{n-1}) e^{i k_0 x_{n-1}}  \nonumber\\
&\times& \ldots d_v (x_2 -x_1) a^{\dagger} (x_1 ) e^{i k_0 x_1} d_v (x_1 -x),
\nonumber
\eeq
\beq
S_v^c [y,x] &=& \tilde{\tilde{d}}_v (y-x)  + \sum_{n=1}^{\infty} \lambda^n \int  {\cal D} x_{n} \label{Stc} \\
& \times & \tilde{d}_v (y - x_n) a^{\dagger} (x_n) e^{i k_0 x_n} \nonumber\\
&\times& d_v (x_n - x_{n-1}) a^{\dagger} (x_{n-1})e^{i k_0 x_{n-1}}  \nonumber\\
&\times & \ldots d_v (x_2 -x_1) a^{\dagger} (x_1 ) e^{i k_0 x_1} \tilde{d}_v (x_1 -x), \nonumber
\eeq
\beq
S_v^{\bar{a}} [y,x]  &=& \tilde{d}_v (y-x) + \sum_{n=1}^{\infty} \lambda^n \int {\cal D} x_{n}  \label{Sta} \\
& \times & \tilde{d}_v (y - x_n) a^{\dagger} (x_n) e^{i k_0 x_n} \nonumber\\
&\times& d_v (x_n - x_{n-1}) a^{\dagger} (x_{n-1})e^{i k_0 x_{n-1}}  \nonumber\\
&\times & \ldots d_v (x_2 -x_1) a^{\dagger} (x_1 ) e^{i k_0 x_1} d_v (x_1 -x), 
\nonumber
\eeq
as well as
\beq
S_v &=& D^{\dagger} (v) S_0 D(v) \nonumber \\
&=& S_v^a [L/2, -L/2] + S^b_v [L/2, -L/2] \frac{\lambda}{\sqrt{2 \Gamma}} A_0^{\dagger},
\label{S-factor}
\eeq
where
\beq
d_v (x) &=&  - \frac{p_+ + p_-}{p_+ - p_-} \left[ e^{- i p_+ x} - e^{- i p_- x}\right] , \label{dv}\\
\tilde{d}_v (x) &=&  - \frac{p_-}{p_+ - p_-}  e^{- i p_+ x} + \frac{p_+ }{p_+ - p_-} e^{- i p_- x},
\label{tdv}\\
\tilde{\tilde{d}}_v (x) &=& - \frac{p_-^2}{p_+^2 - p_-^2} e^{- i p_+ x} + \frac{p_+^2}{p_+^2 -p_-^2} e^{- i p_- x} \label{ttdv} \nonumber \\
& \equiv &  \tilde{d}_v (x) + \frac{i \lambda v^*}{(\delta +i \Gamma) \sqrt{L}} d_v (x) 
\eeq
are the dressed single-photon propagators, and $p_{\pm} \equiv p_{\pm} (v^*)= \frac{-(\delta+i \Gamma) \pm \sqrt{(\delta+i \Gamma)^2 + 8 \Gamma \alpha v^*/L}}{2} $.

Additional details on evaluation of \eq{Sa}-\eq{Sta} are presented in the Appendix \ref{dress_det}.

\subsection{Factorization property}

To evaluate the FCS  we prove the following key property of {\it generalized} $m$-point correlation functions: their {\it factorization} into the $(m+1)$-fold product of the two-point functions.

Generalized correlation functions are defined by
\beq
& &G_{\beta' \beta}^{(m)} (\{z_l \}; y,x) \label{corr_func} \\
&=& \langle 0  | S_{u}^{\beta' \, \dagger} [y,x] \left( :  \prod_{l=1}^m a^{\dagger} (z_l) a (z_l) : \right) S_{v}^{\beta} [y,x] |0 \rangle , \nonumber
\eeq
where the ``time-forward'' $S_v^{\beta}$ and the  ``time-backward'' $S_u^{\beta' \, \dagger}$ scattering operators depend on {\it different} and {\it arbitrary} displacement parameters $v$ and $u$, respectively. We also assume here that $y>z_m > \ldots > z_1 > x$. Note that an additional symbol for the path-ordering in this expression is not required: operators $a (z_l)$ and $a (z_{l'})$ commute with each other, since (i) they are bosonic, and (ii) they are written in the interaction picture (which is equivalent to the Schr\"{o}dinger picture in the co-moving frame). 

We observe that the relations \eq{split_alg} are also fulfilled by the dressed operators $S_v^{\beta}$: one has just to dress \eq{split_alg} with $D(v)$. Along with the property $d_v (0)=0$, this implies that all relations \eq{aS0a}-\eq{comp_am} also remain valid under the replacement $S_0^{\beta} \to S_v^{\beta}$. These properties allow us to split both $S_v^{\beta}$ and $S_u^{\beta' \, \dagger}$ into $m+1$ operators defined on the ordered intervals, see Eq.~\eq{comp_am}. Applying the Wick's theorem, we contract operators from $S_v^{\beta}$ and $S_u^{\beta' \, \dagger}$ belonging to the same interval; in total we have $m+1$ pairwise contractions of intervals. This procedure leads to a factorization of \eq{corr_func}
\beq
& & G_{\beta' \beta}^{(m)} (\{z_l\};y,x) 
= |\lambda|^{2m}  \mathcal{G}_{\sigma (\beta') \sigma (\beta)} (y-z_m) \nonumber\\ 
& & \qquad \times\left[ \prod_{l=1}^{m-1}  \mathcal{G}_{bb} (z_{l+1} -z_l) \right] \mathcal{G}_{\mu (\beta') \mu (\beta)} (z_1 - x)
\label{fact_prop}
\eeq
into the product of $m+1$  two-point functions defined by
\beq
\mathcal{G}_{\beta' \beta} (z_{l+1}-z_l) &=&  \langle 0 | S_u^{\beta' \, \dagger} [z_{l+1},z_l] S_v^{\beta} [z_{l+1},z_l]| 0\rangle .
\label{G_def}
\eeq

In the following we will also use the special case of \eq{G_def} with $u=v=\alpha$
\beq
\bar{\mathcal{G}}_{\beta' \beta} (z_{l+1}-z_l) &=&  \langle 0 | S_{\alpha}^{\beta' \, \dagger} [z_{l+1},z_l] S_{\alpha}^{\beta} [z_{l+1},z_l]| 0\rangle ,
\label{G_def0}
\eeq
corresponding to the standard definition of the correlation functions. The factorization property \eq{fact_prop} in this case is well-known (see, e.g., Ref.~\cite{MaWo}).

\section{Computation of the FCS} 
\subsection{Detailing the definition}

In our setup shown in Fig.~\ref{fig:fcs} we have the right-moving photons in the incoming state. Therefore, in the outgoing state the right-moving photons correspond to the transmitted particles, while the left-moving photons are those which are reflected. Extending the definition \eq{Fdef} we represent
\beq 
F_{r/l,\tau} (\chi_0 ) &=& \langle 0 | D_o^{\dagger} (\alpha) D_e^{\dagger} (\alpha)  S_0^{(e) \, \dagger} \nonumber \\
 & & \times e^{i \chi_0 N_{r/l, \tau} } S_0^{(e)} D_e (\alpha) D_o (\alpha)| 0\rangle , 
\label{Fdef2}
\eeq
where
\beq
N_{r/l,\tau} = \int_{z_1}^{z_2} d x \frac{a_e^{\dagger} (x) \pm a_o^{\dagger} (x)}{\sqrt{2}} \frac{a_e (x) \pm a_o (x)}{\sqrt{2}} .
\eeq
Since $D_o (\alpha)$ commutes with  $D_e (\alpha)$ and  $S_0^{(e)}$, 
and it holds
\beq
D_o (\alpha)^{\dagger} N_{r/l,\tau} D_o (\alpha) = D_e (\pm \alpha)^{\dagger} N_{r/l,\tau} D_e (\pm\alpha),
\eeq
we can cast
\eq{Fdef2} to
\beq
F_{r/l,\tau} (\chi ) &=& \langle 0 |   D_e^{\dagger} (\alpha)  S_0^{(e) \, \dagger} D_e^{\dagger} (\pm \alpha) \nonumber \\
 & & \times e^{i \chi_0 N_{r/l, \tau} } D_e (\pm \alpha) S_0^{(e)} D_e (\alpha) | 0\rangle .
\label{Fdef3}
\eeq
Using the identity 
\beq
& & e^{i \chi_0 N_{r/l,\tau}} = : e^{(z_0-1) N_{r/l,\tau}} : \label{id_norm1} \\
&=& 1 + \sum_{m=1}^{\infty} (z_0-1)^m \int {\cal D}z'_{m}\left( :  \prod_{k=1}^m a_{r/l}^{\dagger} (z'_k) a_{r/l} (z'_k) : \right),\nonumber
\eeq
where $z_0 = e^{i \chi_0}$, and taking into account that in \eq{Fdef3} there are only even operators from both sides of $e^{i \chi_0 N_{r/l,\tau}}$ as well as the vacuum average is performed, we can integrate out the odd modes. It can be effectively done by replacing  $a_{r/l}^{\dagger} (z'_k) \to \frac{a_e^{\dagger} (z'_k)}{\sqrt{2}}$ and $a_{r/l} (z'_k) \to \frac{a_e (z'_k)}{\sqrt{2}}$ in \eq{id_norm1}. This eventually results in the replacement
\beq
e^{i \chi_0 N_{r/l,\tau}} \to e^{i \chi N_{e, \tau}},
\eeq 
where
\beq
N_{e,\tau} = \int_{z_1}^{z_2} d x a_e^{\dagger} (x) a_e (x),
\eeq
and  $\chi$ is defined by $z-1 = \frac{z_0 -1}{2}$ and $z=e^{i \chi}$. Thus,
\beq
F_{r/l,\tau} (\chi ) &=& \langle 0 |   D_e^{\dagger} (\alpha)  S_0^{(e) \, \dagger} D_e^{\dagger} (\pm \alpha) \nonumber \\
 & & \times e^{i \chi N_{e, \tau} } D_e (\pm \alpha) S_0^{(e)} D_e (\alpha) | 0\rangle .
\label{Fdef3a}
\eeq

For the chiral model, we define the FCS by
\beq
F_{e,\tau} (\chi) = \langle 0 |   D_e^{\dagger} (\alpha)  S_0^{(e)\, \dagger} e^{i \chi N_{e,\tau}} S_0^{(e)} D_e (\alpha) | 0\rangle .
\label{Fdef4}
\eeq
This expression is very similar to \eq{Fdef3a}, so we can combine them together
\beq
F_{\tau}^{(\kappa)} (\chi) &=& \langle 0 |   D^{\dagger} (\alpha)  S_0^{\dagger} D^{\dagger} ((\kappa -1) \alpha) \nonumber \\
 & & \times e^{i \chi N_{\tau} } D ((\kappa -1) \alpha) S_0 D (\alpha) | 0\rangle ,  
\label{Fdef5}
\eeq
where the parameter $\kappa =0,1,2$ distinguishes between the reflected, chiral, and transmitted modes, respectively. We have also suppressed the label $e$, since all subsequent calculations will be performed in the effective chiral basis using the results of the previous section. 

After a simple transformation of \eq{Fdef5} we find
\beq
F_{\tau}^{(\kappa)} (\chi) &=&  e^{-\kappa^2 |\alpha |^2} \langle 0 | S_0^{\dagger} e^{- (\kappa -1) \alpha b_{k_0}^{\dagger}} e^{\kappa \alpha^* b_{k_0}} 
\nonumber\\
& & \times  e^{i \chi N_{\tau}}  e^{\kappa \alpha b_{k_0}^{\dagger}} e^{- (\kappa-1) \alpha^* b_{k_0}} S_0 |0 \rangle ,
\label{fcs_def}
\eeq
which is our starting expression for a computation of the FCS.

\subsection{Integrating out the ``future'' and the ``past''}
\label{int_out}

The operator $N_{\tau} = \int_{z_1}^{z_2} dx a^{\dagger} (x) a (x)$ in the definition \eq{fcs_def} contains only fields belonging to the counting interval $[z_1,z_2]$. In turn, the operator
\beq
b_{k_0} = \frac{1}{\sqrt{L}} \int_{-L/2}^{L/2} d x \, a (x) \, e^{-i k_0 x} = \bar{b}_{k_0} + \tilde{b}_{k_0}
\eeq
contains the counting part 
\beq
\bar{b}_{k_0} &=& \frac{1}{\sqrt{L}} \int_{z_1}^{z_2} d x \, a (x) \, e^{-i k_0 x} ,
\eeq
and its complement $\tilde{b}_{k_0} = b_{k_0}^f + b_{k_0}^p$. The latter consists of the ``future'' ($z_1 >x>-L/2$)  and ``past'' ($L/2 > x >z_2$) parts, see Fig.~\ref{fig:fcs}, which are defined by
\beq
b^f_{k_0} &=& \frac{1}{\sqrt{L}} \int_{-L/2}^{z_1} d x \, a (x) \, e^{-i k_0 x} ,\\
b^p_{k_0} &=& \frac{1}{\sqrt{L}} \int_{z_2}^{L/2} d x \, a (x) \, e^{-i k_0 x}.  
\eeq

Integrating out the states lying outside of the counting interval (see the Appendix \ref{deriv_fcs}), we obtain
\beq
& &  F_{\tau}^{(\kappa)} (\chi) e^{-(z-1) \kappa^2 |\alpha |^2 \frac{\tau}{L}} \nonumber \\
&=&   \left[ 1 + R (T) \left(1 - \frac{|\lambda |^2}{2 \Gamma} \right) - 2\, \mathrm{Re} \, C (T) \right]   \nonumber \\
& &  \times \left( \Lambda_{aa} (\tau)+ \frac{|\lambda|^2}{2 \Gamma} \Lambda_{bb} (\tau) \right) \nonumber \\ 
&+& R (T) \left( \Lambda_{cc} (\tau)  + \frac{|\lambda|^2}{2 \Gamma}  \Lambda_{\bar{a} \bar{a}} (\tau) \right)  \nonumber \\
&+& [C (T)-R(T)]  \left( \Lambda_{ac} (\tau)  + \frac{|\lambda|^2}{2 \Gamma} \Lambda_{b \bar{a}} (\tau) \right) \nonumber \\
&+&  [C^* (T)- R (T)] \left( \Lambda_{ca} (\tau)  + \frac{|\lambda|^2}{2 \Gamma}  \Lambda_{\bar{a} b} (\tau) \right) ,
\label{fcs:c}
\eeq
where 
\beq
\Lambda_{\beta' \beta} (\tau) = \langle 0|   S_{u}^{\beta' \, \dagger} [z_2,z_1]  e^{i \chi N_{\tau}} S_{v}^{\beta} [z_2,z_1] | 0 \rangle
\label{Lam_term}
\eeq
evaluated at $u = z_{\kappa} \alpha$ and $v^* = z_{\kappa} \alpha^*$, where $z_{\kappa} = \kappa (z - 1) +1$, and
\beq
R (T) &=& \bar{\mathcal{G}}_{bb} (L/2 -z_2) , \\
C (T) &=& \bar{\mathcal{G}}_{ab} (L/2-z_2)  .
\eeq
The properties of the functions $R (T)$ and $C (T)$ are summarized in the Appendix \ref{app:rcmn}.

The last remaining step is to compute the terms \eq{Lam_term}. 
Using the identity 
\beq
& & e^{i \chi N_{\tau}} = : e^{(z-1) N_{\tau}} : \\
& & =1 + \sum_{m=1}^{\infty} (z-1)^m \int {\cal D}z'_{m}\left( :  \prod_{l=1}^m a^{\dagger} (z'_l) a (z'_l) : \right),\nonumber
\eeq
the definition of the generalized correlation functions \eq{corr_func}, and their factorization property \eq{fact_prop}, we express
\beq
& & \Lambda_{\beta' \beta} (\tau) = \mathcal{G}_{\beta' \beta} (z_2 -z_1) + \sum_{m=1}^{\infty} \left [(z-1) |\lambda|^{2} \right]^m  \nonumber \\
& & \qquad \times \int {\cal D}z'_{m} \nonumber 
\mathcal{G}_{\sigma (\beta') \sigma (\beta)} (z_2-z'_m) \nonumber\\ 
& & \qquad \times\left[ \prod_{l=1}^{m-1}  \mathcal{G}_{bb} (z'_{l+1} -z'_l) \right] \mathcal{G}_{\mu (\beta') \mu (\beta)} (z'_1 - z_1).
\eeq
Defining the Laplace transform for $x>0$ and its inverse
\beq
\mathcal{G}_{\beta' \beta} (p) &=& \int_0^{\infty} d x  \, \mathcal{G}_{\beta' \beta} (x) e^{i p x}, \label{LT} \\
\mathcal{G}_{\beta' \beta} (x) &=& \int_{-\infty +i 0}^{\infty +i 0} \frac{dp}{2 \pi} e^{-i p x}  \mathcal{G}_{\beta' \beta} (p) ,
\eeq
we establish
\beq
& & \Lambda_{\beta' \beta} (\tau) =  \int \frac{d p}{2 \pi} e^{- i p \tau} \left[ \mathcal{G}_{\beta' \beta} (p) \right. \nonumber \\
& & \left. +  (z-1) |\lambda |^2 \frac{\mathcal{G}_{\sigma (\beta') \sigma (\beta)} (p) \mathcal{G}_{\mu (\beta') \mu (\beta)} (p)}{1 - (z-1) |\lambda|^2 \mathcal{G}_{bb} (p)} \right]. 
\label{lam_fin}
\eeq

The Laplace transforms \eq{LT} for all components $\mathcal{G}_{\beta' \beta} (p)$ are listed in the Appendix \ref{LT_app}. With their help we find (see the Appendix \ref{app:norm}) the following expressions
\beq
& & \Lambda_{aa} (\tau) + \frac{|\lambda|^2}{2 \Gamma} \Lambda_{bb} (\tau)  \label{lam_aa_res} \\
&=& \int \frac{d p}{2 \pi}   \frac{ i e^{-i p \tau} }
{p  +  \frac{\Omega_r^2}{R_0 (p)} (z-1) (p+i \Gamma) [ i \Gamma -  \kappa  (p+i \Gamma)]}  \nonumber \\
& & \times  \left[1 - \frac{\Omega_r^2 \kappa (z-1) (p+i \Gamma)}{2 R_0 (p)}  \right], \nonumber \\
& & \Lambda_{cc} (\tau)  + \frac{|\lambda|^2}{2 \Gamma}  \Lambda_{\bar{a} \bar{a}} (\tau) \label{lam_cc_res}\\
&=& \int \frac{d p}{2 \pi}  \frac{i e^{-i p \tau} }{p  +  \frac{\Omega_r^2}{R_0 (p)} (z-1) (p+i \Gamma) [ i \Gamma -  \kappa  (p+i \Gamma)]}\nonumber \\
& & \times  \left[ 1+\frac{|\lambda|^2}{2 \Gamma}  +\frac{i \Gamma \Omega_r^2 (z-1) (1-\kappa)}{2 R_0 (p)} \left( 1 + \frac{(p + 2 i \Gamma)^2}{\delta^2 + \Gamma^2}   \right) \right. \nonumber \\ 
& & \left.  - \frac{\Omega^2_r \kappa (z -1)(p+i \Gamma) }{R_0 (p)}  \left( 1 +  \frac{ \Omega_r^2  (\kappa (z -1) +3)}{ 8 (\delta^2 +\Gamma^2) } \right) \right] , \nonumber \\
& & \Lambda_{ca} (\tau)  + \frac{|\lambda|^2}{2 \Gamma}  \Lambda_{\bar{a} b} (\tau) 
\label{lam_ca_res} \\
&=& \int \frac{d p}{2 \pi}   \frac{ i e^{-i p \tau}}{p  +  \frac{\Omega_r^2}{R_0 (p)} (z-1) (p+i \Gamma) [ i \Gamma -  \kappa  (p+i \Gamma)]} \nonumber \\
& & \times \left[ 1 + \frac{ i \Gamma \Omega_r^2 (z-1)  (1-\kappa)  (p + \delta +  i \Gamma )}{2 R_0 (p) (\delta -i  \Gamma) } \right. \nonumber \\
 & & \left.  \quad -   \frac{\Omega_r^2 \kappa (z - 1)(p+i \Gamma)}{2 R_0 (p)}  \left( 1 + \frac{p(p + \delta+i \Gamma ) }{ 2 (p+i \Gamma) (\delta -i  \Gamma) } \right) \right],  \nonumber
\eeq
where $\Omega_r = \sqrt{8 \Gamma |\alpha|^2/L}$ denotes the Rabi frequency (note also the relation $\frac{|\lambda|^2}{2 \Gamma} = \frac{\Omega_r^2}{4 (\delta^2+ \Gamma^2)}$), and
\beq
R_0 (p) &=& p^3 + 4 i \Gamma p^2 - p \left( \Omega_r^2 + \delta^2 + 5 \Gamma^2\right) \nonumber \\
& & -  i \Gamma \left( \Omega_r^2 + 2 \delta^2 +  2 \Gamma^2 \right) 
\label{R0}
\eeq 
is the third-order polynomial. An expression for $ \Lambda_{ac} (\tau) + \frac{|\lambda|^2}{2 \Gamma} \Lambda_{b\bar{a}} (\tau)$ is obtained from \eq{lam_ca_res} by flipping the sign $\delta \to - \delta$.

The expressions \eq{fcs:c} and \eq{lam_aa_res}-\eq{R0} completely define the FCS for reflected, chiral, and transmitted photons in the model under consideration. One should also keep in mind that for $\kappa =0,2$ it is necessary to replace in the end of calculation $z-1 \to \frac{z_0 -1}{2}$, since the counting parameter for the reflected and transmitted photons is $\chi_0$, and it is related to $z_0 = e^{i \chi_0}$ (see also discussion before \eq{Fdef3a}).

\subsection{Normalization of probability distribution}

We must check normalization of the probability distribution generated by \eq{fcs:c}, which is expressed by the condition $F_{\tau}^{(\kappa)} (0)=1$. Noticing that at $\chi=0$
the functions $\Lambda_{\beta' \beta} (\tau)$, $\mathcal{G}_{\beta' \beta} (\tau)$, and $ \bar{\mathcal{G}}_{\beta' \beta} (\tau)$ coincide with each other, we check the following identities 
\beq
\bar{\mathcal{G}}_{aa} (\tau) + \frac{|\lambda |^2}{2 \Gamma} \bar{\mathcal{G}}_{bb} (\tau) &=& 1 , \label{norm_id1} \\
\bar{\mathcal{G}}_{cc} (\tau) + \frac{|\lambda |^2}{2 \Gamma} \bar{\mathcal{G}}_{\bar{a}\bar{a}} (\tau) &=& 1 + \frac{|\lambda |^2}{2 \Gamma} ,  \label{norm_id2} \\
\bar{\mathcal{G}}_{ca} (\tau) + \frac{|\lambda |^2}{2 \Gamma} \bar{\mathcal{G}}_{\bar{a}b} (\tau) &=& 1 , \label{norm_id3}
\eeq
which hold for arbitrary $\tau$, by setting $z=1$ in \eq{lam_aa_res}-\eq{lam_ca_res}, and insert them into \eq{fcs:c}. We see that $F_{\tau}^{(\kappa)} (0) =1$ is indeed fulfilled for all $\tau$ and $T$. This also means that the scattering wavefunction is properly normalized.

\subsection{Limiting cases of the waiting time $T$}

\subsubsection{Waiting regime $T \to \infty$}

One of the important detection regimes is when the detector's waiting time is long enough, $T \to \infty$, what physically means that $T$ is much larger than all system's time scales, but still smaller than $L$. Depending on the context we will also call this regime {\it stationary} (for the resonance fluorescence) and {\it bulk} (for the entanglement entropy, see below).

The stationary values $R (\infty) = C (\infty) =\frac{ \Gamma }{\Gamma + |\lambda|^2}$ (see the Appendix \ref{app:rcmn}) allow us to express \eq{fcs:c} as
\beq
& &  F_{\tau}^{(\kappa)} (\chi) e^{-(z-1) \kappa^2 |\alpha |^2 \frac{\tau}{L}}  \label{fcs:inf} \\
&=&   \frac{|\lambda |^2}{2 (\Gamma+|\lambda |^2)} \left( \Lambda_{aa} (\tau)+ \frac{|\lambda|^2}{2 \Gamma} \Lambda_{bb} (\tau) \right) \nonumber \\ 
& & + \frac{\Gamma}{\Gamma+|\lambda |^2} \left( \Lambda_{cc} (\tau)  + \frac{|\lambda|^2}{2 \Gamma}  \Lambda_{\bar{a} \bar{a}} (\tau) \right)  \nonumber \\
&=& \int \frac{d p}{2 \pi}   \frac{ i e^{-i p \tau} W^{(\kappa)} (p)}{p  +  \frac{\Omega_r^2}{R_0 (p)} (z-1) (p+i \Gamma) [ i \Gamma -  \kappa  (p+i \Gamma)]} , \nonumber
\eeq
where
\beq
W^{(\kappa)} (p) &=& 1+ \frac{\Gamma \Omega_r^2 (z-1) }{(|\lambda |^2 + \Gamma ) R_0 (p)}  \label{Wk} \\
& \times&  \left[ \frac{i \Gamma (1-\kappa)}{2} \left( 1 + \frac{(p + 2 i \Gamma)^2}{\delta^2 +\Gamma^2}   \right) \right. \nonumber \\
&-& \left.  \kappa (p+i \Gamma)   \left( 1 +  \frac{ \Omega_r^2  (\kappa (z -1) +4)}{ 8 (\delta^2 +\Gamma^2) } \right) \right].\nonumber
\eeq

To find the relation of this result to the photon number statistics in the stationary resonance fluorescence \cite{Mandel},\cite{Smirnov-Troshin}, we must consider the case $\kappa=0$ corresponding to the reflected photons. We note that the fluorescent photons do not interfere with the driving field, and this is precisely the case for the left-moving photons in the presence of the right-propagating driving field. In fact, we find the full agreement of \eq{fcs:inf}, \eq{Wk} at $\kappa=0$ with the result of Mandel \cite{Mandel} and especially with that of Smirnov and Troshin \cite{Smirnov-Troshin}, who have also expressed it in terms of the Laplace transform.

\subsubsection{Waiting regime $T =0$}

In this regime  (which we will also call {\it boundary} in the context of the entanglement entropy below) the detection starts from the forefront of the pulse. Using the initial values $R(0) =C (0)=0$ (see the Appendix \ref{app:rcmn}) we cast \eq{fcs:c} to 
\beq
F_{\tau}^{(\kappa)} (\chi) e^{-(z-1) \kappa^2 |\alpha |^2 \frac{\tau}{L}} = \Lambda_{aa} (\tau) + \frac{|\lambda|^2}{2 \Gamma} \Lambda_{bb} (\tau) .
\eeq
In the case $\kappa =0$ our result
\beq
F_{\tau}^{(0)} (\chi_0) 
&=& \int \frac{d p}{2 \pi}   \frac{ i e^{-i p \tau} }
{p  +  \frac{i \Gamma \Omega_r^2}{2 R_0 (p)} (z_0 -1) (p+i \Gamma) }
\eeq
identically coincides with that of Lenstra \cite{Lenstra} which was derived for the corresponding regime of the resonance fluorescence.

\subsection{Limit of long counting time $\tau$}

When the counting time $\tau$ is much larger than all system's  time scales, the main contribution to $F_{\tau}^{(\kappa)}$ comes from the pole in the vicinity of zero, the contributions from other poles being exponentially suppressed. The dependence on $(z-1)$ in the numerators of Eqs.~\eq{lam_aa_res}-\eq{lam_ca_res} can be also neglected in the large-$\tau$ limit, and we obtain
\beq
& & F_{\tau}^{(\kappa)} (\chi) e^{-(z-1)\kappa^2 |\alpha |^2 \frac{\tau}{L}}  \label{pre_pois}  \label{fcs:tau} \\
& & \approx  \int \frac{d p}{2 \pi}   \frac{ i e^{-i p \tau} }
{p  +  \frac{\Omega_r^2}{R_0 (p)} (z-1) (p+i \Gamma) [ i \Gamma -  \kappa  (p+i \Gamma)]} ,
 \nonumber
\eeq
independently of the waiting time $T$.

Setting $p=0$ in the remaining term $\propto (z-1)$ in the denominator of \eq{fcs:tau}, we arrive at the Poissonian distribution $F_{\tau}^{(\kappa)} (\chi) \approx e^{(z-1) \langle N \rangle}$ characterized by the mean value
\beq
\langle N \rangle = \tau \left[ \kappa^2  \frac{|\alpha |^2}{L} +  \frac{\Omega_r^2 \Gamma ( 1-  \kappa )}{\Omega_r^2 + 2 \delta^2 + 2 \Gamma^2}  \right].
\label{pois_mean}
\eeq
In particular, in the chiral model the mean number of photons remains the same, $\tau |\alpha|^2/L$, as in the incident beam, while the mean numbers of reflected and transmitted photons are
\beq
\langle N_l \rangle &=& \frac12 \langle N \rangle_{\kappa=0} = \frac{\tau |\alpha_0|^2}{ L}  \frac{ \Gamma^2}{\frac12 \Omega_r^2 + \delta^2 + \Gamma^2} , \label{meanNl} \\
\langle N_r \rangle &=& \frac12 \langle N \rangle_{\kappa=2} = \frac{\tau |\alpha_0|^2}{L}  \frac{\frac12 \Omega_r^2 + \delta^2 }{\frac12 \Omega_r^2 + \delta^2 + \Gamma^2}. \label{meanNr}
\eeq
Recall the necessary replacement $z-1 \to \frac{z_0 -1}{2}$, giving the additional factor $1/2$ in the two last expressions.

To find corrections to the Poissonian distribution, we expand the denominator in \eq{pre_pois} to
the linear order in $p$
\beq
& \approx &  [1+ (z-1) Z] \left( p - i \frac{(z-1)}{1+ (z-1) Z}  \frac{\Omega_r^2 \Gamma (1-\kappa)}{\Omega_r^2 + 2 \delta^2  +2 \Gamma^2} \right) \nonumber \\
& \approx &   p - i (z-1)  (1-(z-1) Z) \frac{\Omega_r^2 \Gamma  (1-\kappa)}{\Omega_r^2 + 2 \delta^2  +2 \Gamma^2},
\eeq
where
\beq
Z &=& \frac{d}{dp} \left[ \frac{\Omega_r^2}{R_0 (p)} (p+i \Gamma) ( i \Gamma - \kappa (p+i \Gamma))\right]_{p=0} \label{Zfactor} \\
&=& \Omega_r^2 \frac{\kappa (\Omega_r^2 + 3 \delta^2 - \Gamma^2) +3 \Gamma^2-  \delta^2 }{(\Omega_r^2 + 2 \delta^2 + 2 \Gamma^2)^2}  .\nonumber
\eeq
Thereby we achieved the $O ((z-1)^2)$ correction to the pole position in the vicinity of zero, which leads to a modification of the Poissonian form of the generating function
\beq
F_{\tau}^{(\kappa)} (\chi) \approx e^{(z-1) \langle N \rangle} e^{-\tau (z-1)^2 Z  \frac{\Omega_r^2 \Gamma (1-\kappa)}{\Omega_r^2 + 2 \delta^2  +2 \Gamma^2} }.
\label{post_pois}
\eeq

\begin{figure}[t]
    \includegraphics[width=0.48\textwidth]{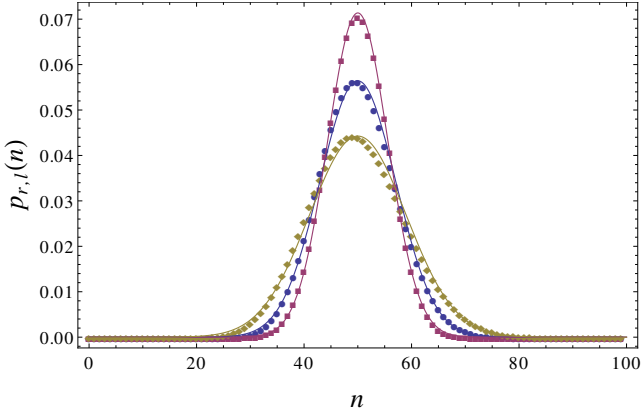}
    \caption{(Color online) Probability distributions $p_r (n)$ (dark yellow diamonds) and $p_l (n)$ (magenta squares) calculated on the basis of Eq.~\eq{post_pois} for $\delta=0$, $\Omega_r = \sqrt{2}\Gamma$, and $\langle N_r \rangle = \langle N_l \rangle = \frac{\Gamma \tau}{4} =50$. Blue circles indicate the Poissonian distribution with the same mean value. Solid lines correspond to the Gaussian approximations $\frac{e^{-\frac{(n -\langle N_r \rangle )^2}{2\langle N_r \rangle (1+Q_r) }}}{\sqrt{2 \pi \langle N_r \rangle (1+Q_r)}}$ (dark yellow), $\frac{e^{-\frac{(n -\langle N_r \rangle )^2}{2\langle N_r \rangle }}}{\sqrt{2 \pi \langle N_r \rangle}} $ (blue), and $\frac{e^{-\frac{(n -\langle N_r \rangle )^2}{2\langle N_r \rangle (1+Q_l) }}}{\sqrt{2 \pi \langle N_r \rangle (1+Q_l)}} $ (magenta), which become rather accurate at large mean values. The Mandel's $Q$-factors, which are $Q_r = \frac58$ and $Q_l =- \frac38$ for the chosen parameters, quantify deviations of the variances from that of the Poissonian distribution.} \label{fig:pn}
\end{figure}

Deviations of \eq{post_pois} from the Poissonian statistics can be quantified in terms of the Mandel's $Q$-factor \cite{Mandel}
\beq
Q = \lim_{\tau \to \infty} \frac{\langle N^{(2)} \rangle -\langle N \rangle^2}{\langle N \rangle}, \label{Qfactor}
\eeq
where $\langle N^{(2)} \rangle$ is the second factorial moment. For $Q<0$ a  distribution is narrower than the Poissonian, and it is called sub-Poissonian; for $Q>0$ a distribution is broader than the Poissonian, and it is called super-Poissonian. If $Q=0$, a distribution is almost indistinguishable from the Poissonian. Performing an expansion of \eq{post_pois} in $(z-1)$ up to the quadratic term we establish
\beq
Q & =& -2 \frac{ Z (1-\kappa) \Gamma^2}{\frac{\kappa^2}{8} (\Omega_r^2 + 2 \delta^2  +2 \Gamma^2)+  \Gamma^2  (1-\kappa)} .
\eeq

We note that in the chiral model ($\kappa=1$) the Mandel's $Q$-factor identically vanishes, rendering it Poissonian. Considering
$\kappa=0$ and $\kappa=2$, we find the factors $Q_l$ and $Q_r$ for the reflected and transmitted photons (recall again the necessary replacement $z-1 \to \frac{z_0 -1}{2}$)
\beq
Q_l &=& \frac12 Q_{\kappa=0}= -  \Omega_r^2 \frac{3 \Gamma^2-  \delta^2 }{(\Omega_r^2 + 2 \delta^2 + 2 \Gamma^2)^2} , 
\label{Ql} \\
Q_r &=& \frac12 Q_{\kappa=2} = \frac{ \Omega_r^2 \Gamma^2}{\frac{1}{2} \Omega_r^2 +  \delta^2  } \frac{2 \Omega_r^2 + 5 \delta^2 +\Gamma^2}{(\Omega_r^2 + 2 \delta^2 + 2 \Gamma^2)^2}.
\label{Qr}
\eeq
We see that for $\delta < \Gamma \sqrt{3}$ the statistics of the reflected photons is sub-Poissonian, and it turns into super-Poissonian for $\delta > \Gamma \sqrt{3}$, while the statistics of the transmitted photons is always super-Poissonian. At $\delta=0$ the expression \eq{Ql} coincides with the original result of Mandel \cite{Mandel} for the photon number statistics in the stationary resonance fluorescence.

In Fig.~\ref{fig:pn} we show the probability distributions $p_r (n)$ and $p_l (n)$ of transmitted and reflected photons for the long counting time $\tau = \frac{200}{\Gamma}$, which are generated by \eq{post_pois}. We also choose $\delta=0$ and $\Omega_r = \sqrt{2} \Gamma$, which provide $\langle N_r \rangle =\langle N_l \rangle$, to ease a comparison of these two distributions. This plot elucidates the physical meaning of the Mandel's $Q$-factor.

We also mention that these theoretical results on photon statistics are supported by the recent measurements of the second-order correlation function \cite{chal1} showing photon antibunching in the reflected field and superbunching in the transmitted field.

\section{Transmission, reflection, and the Mollow triplet}

The mean field and the resonance fluorescence power spectrum, also known as the Mollow triplet \cite{Mollow}, are usually computed using the equation of motion method and the quantum regression theorem. In this section we demonstrate how to obtain these expressions in the framework of our scattering approach.

To compute the mean field $\langle a (z_1) \rangle$ and the first order correlation function $g^{(1)} (z'_1, z_1) = \langle a^{\dagger} (z'_1) a (z_1)  \rangle$ we employ the expression \eq{Fdef5}, in which we replace $e^{i \chi N_{\tau}}$ by $a (z_1)$ and $a^{\dagger} (z'_1) a (z_1)$. Thus,
\beq
\langle a (z_1) \rangle &=&  \langle 0 |   S_{\alpha}^{\dagger} D^{\dagger} (\kappa  \alpha ) a (z_1)D (\kappa \alpha) S_{\alpha} | 0\rangle  \nonumber \\
&=& \frac{\kappa \alpha}{\sqrt{L}} e^{i k_0 z_1} + \langle 0 |   S_{\alpha}^{\dagger} a (z_1) S_{\alpha} | 0\rangle , \label{mf1} \\
\langle a^{\dagger} (z'_1) a (z_1) \rangle &=&  \langle 0 |   S_{\alpha}^{\dagger} D^{\dagger} (\kappa  \alpha ) a^{\dagger} (z'_1) a (z_1)D (\kappa \alpha) S_{\alpha} | 0\rangle  \nonumber \\
&=&  \frac{\kappa^2 |\alpha |^2}{L} e^{-i k_0 (z'_1 - z_1)} 
\nonumber \\
 &+&  \frac{\kappa \alpha^*}{\sqrt{L}} e^{-i k_0 z'_1} \langle  a (z_1) \rangle + \langle  a (z'_1) \rangle^*  \frac{\kappa \alpha}{\sqrt{L}} e^{i k_0 z_1}
 \nonumber \\
 &+&  \langle 0 |   S_{\alpha}^{\dagger} a^{\dagger} (z'_1) a (z_1) S_{\alpha} | 0\rangle .
 \label{g1_1}
\eeq
Note that to find the reflected and transmitted ($\kappa=0,2$) fields and their $g^{(1)}$ functions, it is necessary to multiply  \eq{mf1} and \eq{g1_1} additionally by the factors $1/\sqrt{2}$ and $1/2$.
 
So we see that it suffices to compute 
\beq
 & & \langle 0 | S_{\alpha}^{\dagger} a (z_1) S_{\alpha} |0 \rangle 
\nonumber \\
&=& \langle 0 | S_{\alpha}^{a \, \dagger} [L/2,-L/2] a (z_1) S_{\alpha}^a [L/2,-L/2] |0 \rangle \nonumber\\
&+& \frac{|\lambda|^2}{2 \Gamma} \langle 0 | S_{\alpha}^{b \, \dagger} [L/2,-L/2] a (z_1) S_{\alpha}^b [L/2,-L/2] |0 \rangle \nonumber
\eeq
 and
\beq 
& &\tilde{g}^{(1)} (z'_1 , z_1)  \equiv  \langle 0 | S_{\alpha}^{\dagger} a^{\dagger} (z'_1) a (z_1) S_{\alpha} |0 \rangle 
\nonumber \\
&=& \langle 0 | S_{\alpha}^{a \, \dagger} [L/2,-L/2] a^{\dagger} (z'_1) a (z_1) S_{\alpha}^a [L/2,-L/2] |0 \rangle \nonumber\\
&+& \frac{|\lambda|^2}{2 \Gamma} \langle 0 | S_{\alpha}^{b \, \dagger} [L/2,-L/2] a^{\dagger} (z'_1) a (z_1) S_{\alpha}^b [z_1,-L/2]  |0 \rangle .\nonumber
\eeq

Applying the $\alpha$-shifted versions of \eq{aS0a} and \eq{aS0b} we obtain
\beq
\langle 0 | S_{\alpha}^{\dagger} a (z_1) S_{\alpha} |0 \rangle &=& \lambda e^{i k_0 z_1}
\nonumber \\
& &  \times \left(\langle 0 | S_{\alpha}^{a \, \dagger} [L/2,-L/2] \right. \nonumber \\
& & \qquad \times S_{\alpha}^b [L/2,z_1] S_{\alpha}^a [z_1,-L/2] |0 \rangle 
\nonumber\\
&+& \frac{|\lambda|^2}{2 \Gamma} \langle 0 | S_{\alpha}^{b \, \dagger} [L/2,-L/2] \nonumber \\
& & \left. \qquad \times S_{\alpha}^b [L/2,z_1] S_{\alpha}^b [z_1,-L/2] |0 \rangle \right) \nonumber
\eeq
and
\beq
\tilde{g}^{(1)} (z'_1 , z_1) &=& |\lambda|^2 e^{-i k_0 (z'_1 -z_1)}
\nonumber \\
& & \times \left(\langle 0 | S_{\alpha}^{a \, \dagger} [z'_1,-L/2]  S_{\alpha}^{b \, \dagger} [L/2,z'_1] \right. \nonumber \\
& & \qquad \times S_{\alpha}^b [L/2,z_1] S_{\alpha}^a [z_1,-L/2] |0 \rangle  \nonumber\\
&+& \frac{|\lambda|^2}{2 \Gamma} \langle 0 | S_{\alpha}^{b \, \dagger} [z'_1,-L/2] S_{\alpha}^{b \, \dagger} [L/2,z'_1]    \nonumber \\
& & \left.  \qquad \times S_{\alpha}^b [L/2,z_1] S_{\alpha}^b [z_1,-L/2]  |0 \rangle  \right).\nonumber
\eeq

Next, using \eq{split_alg} we split in the first case the operators $S_{\alpha}^{a,b \, \dagger} [L/2,-L/2]$ into the subintervals $[-L/2, z_1]$ and $[z_1, L/2]$. Assuming $z'_1 > z_1$ in the second case, we split the operators $S_{\alpha}^{a,b \, \dagger} [z'_1,-L/2]$ into the subintervals $[-L/2, z_1]$ and $[z_1, z'_1]$ and the operator $S_{\alpha}^b [L/2,z_1]$ into the subintervals $[z_1, z'_1]$ and $[z'_1,L/2]$. After that, we apply the Wick's theorem, use the identities \eq{norm_id1}-\eq{norm_id3}, and obtain
\beq
& & \langle 0 | S_{\alpha}^{\dagger} a (z_1) S_{\alpha} |0 \rangle = \lambda e^{i k_0 z_1} C (L/2 -z_1), \\
& & \tilde{g}^{(1)} (z'_1 , z_1)  =  |\lambda|^2 e^{-i k_0 (z'_1 -z_1)}  \label{til_g1} \\
& \times & \left(R (T)  M (z'_1 -z_1)  + [C^* (T) -R (T)]   C (z'_1 -z_1)   \right),
\nonumber
\eeq
where $T=L/2 -z'_1$ and $M (\tau) = \bar{\mathcal{G}}_{a \bar{a}} (\tau)$. The properties of the latter function are studied in the Appendix \ref{app:rcmn}.

In the stationary regime $L/2 - z_1 \to \infty$ the mean field equals
\beq
 \langle  a (z_1)  \rangle \approx e^{i k_0 z_1} \left( \frac{\kappa \alpha}{\sqrt{L}} + \frac{\lambda \Gamma}{\Gamma+ |\lambda |^2}\right).
\eeq
In particular, we find the mean reflected ($\kappa=0$) and transmitted ($\kappa=2$) fields (recall also about the additional factor $1/\sqrt{2}$)
\beq
 \langle  a_l (z_1) \rangle & \approx & - \frac{\alpha_0 e^{i k_0 z_1}}{\sqrt{L}}  \frac{i \Gamma (\delta - i \Gamma)}{\frac12 \Omega_r^2 + \delta^2 + \Gamma^2}  , \\
 \langle  a_r (z_1)  \rangle & \approx & \frac{\alpha_0 e^{i k_0 z_1}}{\sqrt{L}}  \left( 1 - \frac{ i \Gamma  (\delta - i \Gamma)}{\frac12 \Omega_r^2 +\delta^2 + \Gamma^2 } \right).
 \eeq
Dividing these expressions by $\frac{\alpha_0 e^{i k_0 z_1}}{\sqrt{L}}$, we obtain the reflection and transmission amplitudes, in full agreement with \cite{coherent1} and \cite{coherent2}. 

We note that for the strong drive the quantities $ |\langle  a_{l/r} (z_1) \rangle |^2$ differ from the mean numbers of photons per unit time $N_{l/r}/\tau$, defined by \eq{meanNl}, \eq{meanNr}. These observables begin to coincide in the limit of the weak driving field $\alpha_0 \to 0$ and $\Omega_r \to 0$, both converging to the single-photon reflection $\frac{\Gamma^2}{\delta^2 + \Gamma^2} \frac{|\alpha_0 |^2}{L}$ and transmission $\frac{\delta^2}{\delta^2 + \Gamma^2} \frac{|\alpha_0 |^2}{L}$ probabilities (times the incident photon density), what indicates the suppression of the inelastic scattering processes.

The information about the inelastic -- Mollow -- part of the power spectrum is contained in $\tilde{g}^{(1)} (z'_1, z_1)$ expressed by \eq{til_g1}, and, more precisely, in the functions
\beq
M (\tau)  &=& \int \frac{d p}{2 \pi} e^{-i p \tau} \left[ \tilde{r} (p) + \frac{|\lambda |^2 \bar{c} (p) c (p)}{1 -|\lambda |^2 r (p)} \right]  \\
&=&  \frac{\Gamma}{\Gamma + |\lambda|^2} \left(1 +\frac{|\lambda|^2}{\Gamma} \int \frac{d p}{2 \pi}  e^{-i p \tau} \frac{i M_0 (p)}{R_0 (p)} \right), \nonumber \\
C (\tau) &=& \int \frac{dp}{2 \pi}  e^{-i p \tau} \frac{c (p)}{1-|\lambda|^2 r (p)} \\
 &=&  \frac{\Gamma}{\Gamma + |\lambda|^2} \left(1 - \int \frac{dp }{2 \pi}  e^{-i p \tau} \frac{i C_0 (p)}{R_0 (p)} \right), \nonumber 
\eeq
in the form of terms containing the third-order polynomial $R_0 (p)$ which is defined in \eq{R0}. Here $r (p) = r (p; \alpha, \alpha^*)$, $\tilde{r} (p) = \tilde{r} (p; \alpha, \alpha^*)$, $c (p) = c (p; \alpha, \alpha^*)$ are the special cases of the functions \eq{r_expl}, \eq{rt_expl}, \eq{c_expl}, and
\beq
& & M_0 (p) =  (p+2 i \Gamma)^2 - \frac{\Omega_r^2}{2}, \label{M0} \\
& & C_0 (p) = M_0 (p) - \frac{ \Gamma+ |\lambda|^2}{\Gamma} (\delta + i \Gamma)  (p+2 i \Gamma) . 
\label{C0}
\eeq
We note the inelastic power spectrum is the same (up to the factor $1/2$) for the chiral and reflected/transmitted photons, therefore it is sufficient to consider only the chiral case.

We analyze first the stationary regime $T \to  \infty$ of \eq{til_g1}. Here we have $R (\infty) = C (\infty) =\frac{\Gamma}{\Gamma + |\lambda|^2}$, and the Mollow spectrum is completely defined by $\frac{\Gamma}{\Gamma + |\lambda|^2}\mathrm{Re} \frac{i M_0 (p)}{R_0 (p)}$, in the full agreement with the known results \cite{Mollow}, \cite{KM-rev}. The roots of $R_0 (p)$ define the positions and widths of all three Mollow peaks, while the functions $M_0 (p)$ [Eq.~\eq{M0}] contribute to their weights.

The expression \eq{til_g1} shows how the shape of the Mollow spectrum evolves in time $T$ toward its stationary value discussed above. At $T=0$ we have $R(0)=C(0)=0$, and the inelastic power spectrum is absent. For finite $T>0$ its weight starts to grow, it acquires the three-peak form, however its transient shape differs from the stationary one due to the presence of the additional contribution in $C_0 (p)$  [Eq.~\eq{C0}] which is not proportional to $M_0 (p)$.

\begin{figure}[t]
    \includegraphics[width=0.48\textwidth]{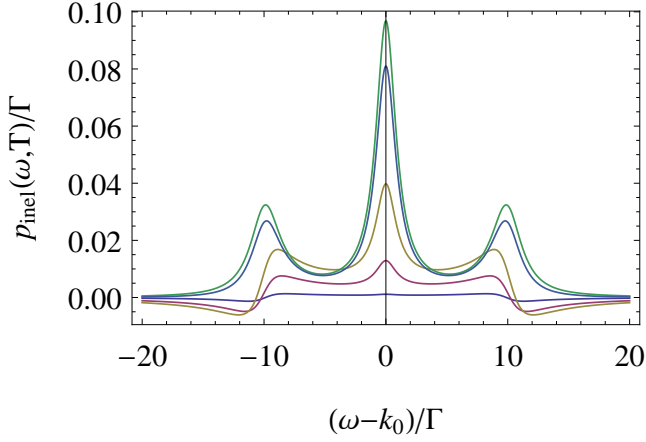}
    \caption{(Color online) Evolution of the Mollow triplet (from bottom to top) with increasing time $T=\frac{0.01}{\Gamma},\frac{0.05}{\Gamma},\frac{0.1}{\Gamma},\frac{1}{\Gamma},\frac{10}{\Gamma}$ for the parameters $\Omega_{r}=10 \Gamma$ and $\delta=0$. For $T=\frac{10}{\Gamma}$ (top green curve) it is already indistinguishable from the stationary shape given by Eq.~\eq{mollow_stat}.} \label{fig:mollow}
\end{figure}

In the formal expression the power spectrum amounts to
\beq
P (\omega) &=& \frac{k_0}{2 \pi T_0} \int d z'_1 d z_1 \langle a^{\dagger} (z'_1) a (z_1 ) \rangle e^{i \omega (z'_1 -z_1)} \nonumber \\
&=& \frac{k_0}{2 \pi T_0} \mathrm{Re} \int_0^{T_0} d T \int_0^{\infty} d \tau e^{i \omega \tau} \nonumber \\
& & \times \langle a^{\dagger} (L/2-T) a (L/2-T - \tau)\rangle ,  \label{Pom}
\eeq
where $T_0$ is the maximal waiting time. The inelastic part of  \eq{Pom} equals
\beq
P_{\mathrm{inel}} (\omega) &=&  \frac{k_0 |\lambda |^2}{2 \pi T_0} \mathrm{Re} \int_0^{T_0} d T \int_0^{\infty} d \tau e^{i (\omega - k_0) \tau}  \label{Pom_in} \\
& & \times (R (T) M_{\mathrm{inel}} (\tau) + [C^* (T) - R(T)] C_{\mathrm{inel}} (\tau)) , \nonumber
\eeq
where
\beq
M_{\mathrm{inel}} (\tau) &=& \frac{|\lambda|^2}{\Gamma +|\lambda|^2} \int \frac{d p}{2 \pi} e^{-i p \tau} \frac{i M_0 (p)}{R_0 (p)}, \\
C _{\mathrm{inel}} (\tau) &=& - \frac{\Gamma}{\Gamma +|\lambda|^2} \int \frac{d p}{2 \pi} e^{-i p \tau} \frac{i C_0 (p)}{R_0 (p)}.
\eeq
Computing \eq{Pom_in} we obtain
\beq
P_{\mathrm{inel}} (\omega) &=& \frac{k_0}{T_0} \int_0^{T_0} d T p_{\mathrm{inel}} (\omega, T) , \\
p_{\mathrm{inel}} (\omega, T) &=&  \frac{|\lambda |^4 R (T)}{2 \pi (\Gamma +|\lambda|^2)}  \mathrm{Re} \left\{ \frac{i M_{0} (\omega -k_0)}{R_0 (\omega -k_0)}
\right\} \label{p_in} \\
& & -  \mathrm{Re}\frac{|\lambda |^2 \Gamma  [C^* (T) - R(T)] i C_{0} (\omega-k_0)}{2 \pi (\Gamma +|\lambda|^2) R_0 (\omega -k_0)}  . \nonumber
\eeq

For $T_0 \to \infty$ we recover the stationary Mollow spectrum
\beq
P_{\mathrm{inel}}^{\mathrm{stat}} (\omega) &=& k_0 \, p_{\mathrm{inel}} (\omega, \infty) \nonumber \\
&=&  \frac{k_0 |\lambda |^4 \Gamma}{2 \pi (\Gamma +|\lambda|^2)^2}  \mathrm{Re} \left\{ \frac{i M_{0} (\omega -k_0)}{R_0 (\omega -k_0)}
\right\} .
\label{mollow_stat}
\eeq
At large Rabi frequency $\Omega_r \gg \Gamma, \delta$ it acquires the most familiar form
\beq
& & P_{\mathrm{inel}}^{\mathrm{stat}} (\omega) = \frac{k_0 \Gamma}{4 \pi}  \\
& \times &  \left\{ \frac{\Gamma}{(\omega- k_0)^2 +\Gamma^2} + \frac12 \sum_{s =\pm} \frac{\frac{3 \Gamma}{2}}{(\omega -k_0 -s \Omega_r)^2+ \frac{9 \Gamma^2}{4}} \right\}. \nonumber 
\eeq

For finite $T_0$ we show in Fig.~\ref{fig:mollow} the augmentation of $p_{\mathrm{inel}} (\omega, T)$ with increasing time $T$.

\section{Reduced density matrix and entanglement entropy}

Explicit knowledge of the scattering state allows us to determine a reduced density matrix of some spatial interval, which we continue to call a counting interval. It suffices to trace out ``past'' and ``future'' states of the full density matrix $(S_0 |\alpha \rangle) (\langle \alpha |S_0^{\dagger})$ by a procedure similar to that described in the section \ref{int_out}. Knowing the reduced density matrix, whose computation by other methods is questionable, we can study the entanglement entropy in our model.

Let us consider an arbitrary many-body operator $\hat{A}$ in the chiral model which is defined on the counting interval $[z_1 , z_2]$. Its average value in the scattering state reads
\beq
\langle \hat{A} \rangle &=& \langle 0 | D^{\dagger} (\alpha) S_0^{\dagger} \hat{A} S_0  D (\alpha) | 0 \rangle \\
&=&   \langle 0 | S_{\alpha}^{\dagger} D^{\dagger} (\alpha) \hat{A} D (\alpha) S_{\alpha} | 0 \rangle \nonumber \\
&=& \langle 0 | S_{\alpha}^{a \, \dagger} [L/2,-L/2] \hat{A}_{\alpha} S_{\alpha}^a [L/2,-L/2] | 0 \rangle 
\nonumber \\
&+& \frac{|\lambda |^2}{2 \Gamma} \langle 0 | S_{\alpha}^{b \, \dagger} [L/2,-L/2] \hat{A}_{\alpha} S_{\alpha}^b [L/2,-L/2] | 0 \rangle , \nonumber 
\eeq
where $\hat{A}_{\alpha} = D^{\dagger} (\alpha) \hat{A}  D (\alpha) $. Repeating the same steps following Eq.~\eq{fcs:fcp}, we obtain an expression for $\langle \hat{A}_{\alpha} \rangle$ analogous to \eq{fcs:c} -- it is only necessary to set $\kappa=1$ and $z=1$, and to replace $e^{i \chi N_{\tau}}$ by $\hat{A}_{\alpha}$. This implies that $\hat{A}$ can be expressed as a trace $\mathrm{Tr}_{\tau} (\hat{A} \hat{\rho}_{\tau})$ over the states in the spatial interval $\tau=z_2-z_1$ with the reduced density matrix of this interval
\beq
\hat{\rho}_{\tau} &=& \left[ 1 + \left( 1- \frac{|\lambda |^2}{2 \Gamma} \right) R (T) - 2 \mathrm{Re} \, C (T) \right] \nonumber\\
&\times& \left( |\psi^a \rangle \langle \psi^a| + \frac{|\lambda |^2}{2 \Gamma} |\psi^b \rangle \langle \psi^b| \right) \nonumber \\
&+& R (T) \left( |\psi^c \rangle \langle \psi^c | + \frac{|\lambda |^2}{2 \Gamma} |\psi^{\bar{a}} \rangle \langle \psi^{\bar{a}} | \right) 
\nonumber \\
&+& \left[ C (T) - R(T) \right]  \left( |\psi^a \rangle \langle \psi^{c} | + \frac{|\lambda |^2}{2 \Gamma} |\psi^b \rangle \langle \psi^{\bar{a}} | \right) \nonumber \\
&+& \left[ C^* (T) - R(T) \right]  \left( |\psi^{c} \rangle \langle \psi^a| + \frac{|\lambda |^2}{2 \Gamma} |\psi^{\bar{a}} \rangle \langle \psi^b| \right) \nonumber \\
&=& \sum_{\beta,\beta'} \rho_{\beta \beta'} (T) | \psi^{\beta} \rangle \langle \psi^{\beta'} |,
\label{red_matr}
\eeq
where 
\beq
| \psi^{\beta} \rangle = D (\alpha) S^{\beta}_{\alpha} [z_2 , z_1] |0 \rangle 
\label{bas1}
\eeq
are linearly independent many-body states. Thus, it turns out that $\hat{\rho}_{\tau} $ describes the states in the effective four-dimensional Hilbert space spanned by $| \psi^{\beta} \rangle$. The reduced density matrix \eq{red_matr} is characterized by four eigenvalues $\lambda_i$, and the entanglement entropy of the interval $\tau$ with the rest of the pulse is then given by
\beq
\mathcal{S} = - \sum_{i=1}^4 \lambda_i \ln \lambda_i .
\eeq

The basis \eq{bas1} is, however, not orthonormal, and the corresponding Gram matrix $\langle \psi^{\gamma'} | \psi^{\gamma} \rangle$ differs from the identity. Our central observation is that its components coincide with the two-point functions \eq{G_def0}, $\langle \psi^{\gamma'} | \psi^{\gamma} \rangle =\bar{\mathcal{G}}_{\gamma' \gamma} (\tau)$. Therefore, the eigenvalues $\lambda_i$ coincide with the eigenvalues of the $4 \times 4$ matrix $\rho (T) \bar{\mathcal{G}} (\tau)$.

 \begin{figure}[t]
    \includegraphics[width=0.45\textwidth]{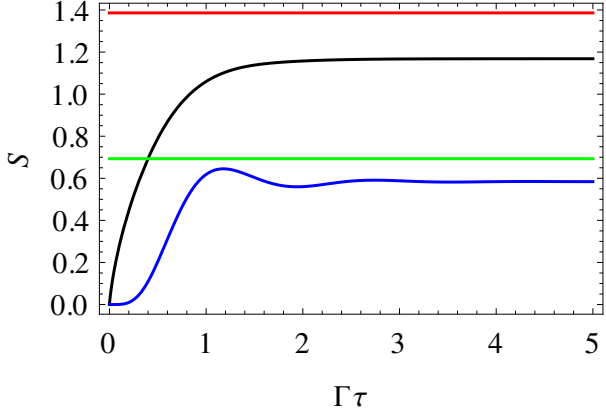}
    \caption{(Color online) Entanglement entropy $\mathcal{S}$ as a function of the subsystem size $\tau$ for $T \to \infty$ (black upper curve) and $T=0$ (blue lower curve); the detuning $\delta=0$ and the Rabi frequency $\Omega_{r}=4 \Gamma$ are the same for both curves.  The horizontal lines indicate the limiting values $\ln 4$ (upper line) and $\ln 2$ (lower line).  } 
    \label{fig:sent1}
\end{figure}

In the {\it bulk} regime $T \to \infty$ we find that one of the eigenvalues of the matrix $\rho (\infty) \bar{\mathcal{G}} (\tau) $ equals
\beq
\lambda_4 (\tau) = - \frac{|\lambda |^4 R (\tau)}{4 \Gamma (\Gamma +|\lambda |^2)}  - |\lambda |^2 \frac{  M (\tau)-1}{2 (\Gamma +|\lambda |^2)}.
\eeq
At small $\tau$ we have
\beq
\lambda_1 (\tau) & \approx & 1 - \frac{|\lambda |^4 \Gamma \tau}{(\Gamma +|\lambda |^2)^2}, \\
\lambda_2 (\tau) & \approx & \frac{|\lambda |^4 \Gamma \tau}{(\Gamma +|\lambda |^2)^2}, \\
\lambda_3 (\tau) & \sim & O (\tau^2), \\
\lambda_4 (\tau) & \sim & O (\tau^3),
\eeq
which leads us  to the following behavior of the entanglement entropy
\beq
\mathcal{S}_{\infty} \approx -  \frac{|\lambda |^4}{(\Gamma +|\lambda |^2)^2} (\Gamma \tau) \ln (\Gamma \tau).
\label{s0_bulk}
\eeq
In the limit of large $\tau \to \infty$ we have
\beq
\lambda_{1,2} = \frac14 \left( 1 \pm \sigma \right)^2 , \quad \lambda_3 = \lambda_4 = \frac{1-\sigma^2}{4},
\eeq
where
\beq
\sigma = \frac{\sqrt{\Gamma (\Gamma + 2 |\lambda |^2)}}{\Gamma + |\lambda |^2} < 1,
\eeq
leading us to the expression
\beq
\mathcal{S}_{\infty}^{\infty} &=& \lim_{\tau \to \infty}  \mathcal{S}_{\infty} \nonumber \\
&=& - (1 +  \sigma ) \ln \frac{1 + \sigma}{2} - (1  -  \sigma) \ln \frac{1 - \sigma}{2} .
\label{s_bulk_stat}
\eeq
For the weak drive $|\lambda|^2 \ll \Gamma$ ($\sigma \to 1$), $\mathcal{S}_{\infty}^{\infty}$ vanishes. It means that in the absence of inelastic processes there are no correlations, and therefore there is no entanglement. For the strong drive $|\lambda|^2 \gg \Gamma$ ($\sigma \to 0$) we obtain $\lambda_i \approx \frac14$, and $\mathcal{S}_{\infty}^{\infty}$ approaches its maximal upper bound $\ln 4$.

\begin{figure}[t]
    \includegraphics[width=0.45\textwidth]{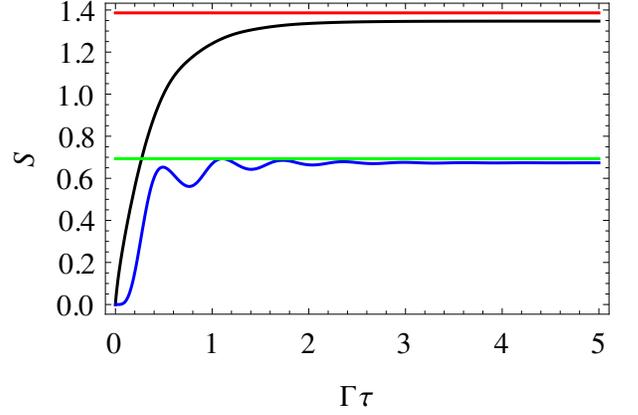}
    \caption{(Color online) The same quantities as in Fig.~\ref{fig:sent1} for the different Rabi frequency $\Omega_{r}=10 \Gamma$.} \label{fig:sent2}
\end{figure}

In the {\it boundary} regime $T=0$ the rank of $\rho (0)$ reduces by two (because of $R (0)=C(0) =0$), and we have only two nonzero eigenvalues
\beq
\lambda_{1,2} (\tau) = \frac12 \left( 1 \pm \sqrt{\left(1- \frac{|\lambda |^2}{\Gamma} R (\tau) \right)^2 - 2 \frac{|\lambda|^2}{\Gamma} |C (\tau)|^2} \right). \nonumber \\
\label{lam12}
\eeq
At small $\tau$ they behave like
\beq
\lambda_1 \approx 1-\frac{(\Omega_r \tau)^2 }{2}, \quad \lambda_2 \approx \frac{(\Omega_r \tau)^2 }{2},
\eeq
yielding
\beq
\mathcal{S}_{0} \approx -  (\Omega_r \tau)^2  \ln (\Omega_r \tau).
\label{s0_bound}
\eeq
At large $\tau$ the eigenvalues \eq{lam12} saturate at the values
\beq
\lambda_{1,2} = \frac{1 \pm \sigma}{2},
\eeq
and therefore
\beq
\mathcal{S}_{0}^{\infty} &=& \lim_{\tau \to \infty}  \mathcal{S}_{0} \nonumber \\
&=& - \frac{1 +  \sigma}{2} \ln \frac{1 + \sigma}{2} - \frac{1  -  \sigma}{2} \ln \frac{1 - \sigma}{2} .
\label{s_bound_stat}
\eeq

It is remarkable that
\beq
\mathcal{S}_{0}^{\infty} = \frac12 \mathcal{S}_{\infty}^{\infty},
\label{sbb}
\eeq
which means that the subsystem lying deep in the bulk of the scattered pulse is twice stronger entangled with the rest system than the subsystem at the forefront of the pulse. The existence of the finite values \eq{s_bulk_stat} and \eq{s_bound_stat} in the large-$\tau$ limit tells us that the area law  is asymptotically fulfilled for the large subsystem size. In our 1D geometry, the ``area'' of the subinterval consists either of two points in the bulk case or of a single point in the boundary case, and this difference in the ``area'' measure is accounted by the factor $1/2$ in \eq{sbb}.

At small $\tau$, the expression \eq{s0_bulk} and \eq{s0_bound} contain the logarithmic terms, which means that the entanglement entropy for the small subsystem size violates the volume ($\sim \tau$) law in our model.

In Fig.~\ref{fig:sent1} and \ref{fig:sent2} the $\tau$-dependence of the entanglement entropy for the bulk (black curve) and boundary (blue curve) cases using the values $\Omega_r =4 \Gamma$ (moderate drive) and  $\Omega_r = 10 \Gamma$ (strong drive). The upper limits $\ln 4$ and $\ln 2$ are indicated by the horizontal lines. In both cases we set the detuning $\delta=0$ for simplicity.

We observe that the bulk entanglement entropy is the monotonously  growing function of the subsystem size. In turn, the boundary entanglement entropy exhibits oscillatory behavior before it reaches the saturation value. For the strong driving field both entropies nearly reach the corresponding maximally allowed values $\ln 4$ and $\ln 2$ at large $\tau$.

Our numerical analysis also shows that the entanglement entropies $\mathcal{S} (\tau)$ for finite values of $T$ lie between the blue and the black curves (not shown in Figs.~\ref{fig:sent1} and \ref{fig:sent2}), though not always being bounded by them, but always being bounded by $\ln 4$ from above.

\section{Conclusion}
We have exactly computed the full counting statistics in the fundamental quantum
optical setup -- a finite size pulse of the coherent light propagating in the multi-mode waveguide and interacting
with the two-level system. These results provide a quantitative determination of many-body
correlation effects of photons mediated by their interaction with the emitter. Our analysis takes into account the spatial parameters of the incident pulse as well as the parameters of the detector -- the waiting time $T$ and the counting time $\tau$, in terms of which the FCS is defined and analyzed. We show that the three types of counting statistics for the reflected, transmitted, and chiral photons have  qualitatively different behavior (sub-Poissonian, super-Poissonian, and Poissonian).  We have analyzed the entanglement entropy of a spatial part of the scattered pulse with the rest of it in the chiral model, and observed the fulfillment of the area law for large subsystem size and the violation of the volume law for  small subsystem size, as well as the oscillatory behavior of the entanglement entropy as a function of the subsystem size $\tau$ in the case of the short waiting time $T$. 

We believe that the full characterization of properties of the scattered coherent light presented here will be requested in future theoretical and experimental studies of many-body effects in fundamental models of quantum nanophotonics and extended for more complicated systems.

\section*{Acknowledgements} 
We benefited a lot from discussions with D. Baeriswyl, A. Fedorov, D. Ivanov, G. Johansson, A. Komnik, M. Laakso, G. Morigi, M. Ringel, and M. Wegewijs. Work of V. G.  is part of the D-ITP consortium, a program of the Netherlands Organisation for Scientific Research (NWO) that is funded by the Dutch Ministry of Education, Culture and Science (OCW).  

\appendix

\section{Additional details of the dressing procedure \eq{disp}}
\label{dress_det}

In the expression \eq{Sa}-\eq{Sta} we introduced the dressed propagators $d_v$, $\tilde{d}_v$, and $\tilde{\tilde{d}}_v$, which emerge after the re-organization of the series for $S_v^{\beta} [y,x]$ and are themselves defined by the following series
\beq
d_v (y-x) \!&\!=\!& \!d_0 (y-x) + \sum_{n=1}^{\infty} \left( \frac{\lambda v^*}{\sqrt{L}} \right)^n\!\!\!\int {\cal D}x_{n} d_0 (y-x_n) 
\nonumber\\
&\times&\!\! d_0 (x_n - x_{n-1}) \ldots d_0 (x_2 -x_1) d_0 (x_1 -x),\nonumber\\
\label{dv_def}
\eeq
\beq
\tilde{d}_v (y-x) &\!=\!& 1 + \sum_{n=1}^{\infty} \left( \frac{\lambda v^*}{\sqrt{L}} \right)^n\int {\cal D}x_{n} \nonumber \\
& & \times  d_0 (x_n - x_{n-1}) \ldots d_0 (x_2 -x_1) d_0 (x_1 -x) \nonumber\\
&=& 1 + \sum_{n=1}^{\infty} \left( \frac{\lambda v^*}{\sqrt{L}} \right)^n\int {\cal D}x_{n}\nonumber \\
& & \times d_0 (y-x_n) d_0 (x_n - x_{n-1}) \ldots d_0 (x_2 -x_1), \nonumber \\
\label{tdv_def}
\eeq
\beq
\tilde{\tilde{d}}_v (y-x) &=& 1 + \frac{\lambda v^*}{\sqrt{L}} (y-x) + \sum_{n=2}^{\infty} \left( \frac{\lambda v^*}{\sqrt{L}} \right)^n\int {\cal D}x_{n}\nonumber \\
& & \times d_0 (x_n - x_{n-1}) \ldots d_0 (x_2 -x_1). \label{ttdv_def}
\eeq
These functions obey the following differential equations
\beq
& & d'_v (x) - i (\delta +i \Gamma) [d_v (x) - \tilde{d}_v (x)] =0, \\
& & \tilde{d}'_v (x) = \frac{\lambda v^*}{\sqrt{L}} d_v (x) , \\
& & \tilde{\tilde{d}}'_v (x) = \frac{\lambda v^*}{\sqrt{L}} \tilde{d}_v (x), \\
& & d_v (0) =0 , \\
& & \tilde{d}_v (0) = \tilde{\tilde{d}}_v (0) =1 .
\eeq
Solving them, we find \eq{dv}-\eq{ttdv}.

\section{Derivation of \eq{fcs:c}}
\label{deriv_fcs}

It is easy to check the following commutation relations
\beq
\, [ \bar{b}_{k_0} , \bar{b}_{k_0}^\dagger ] &=& \frac{\tau}{L} , \\
\, [ \bar{b}_{k_0} , \tilde{b}_{k_0}^\dagger ] &=& 0, \\
\, [ \tilde{b}_{k_0} , \tilde{b}_{k_0}^\dagger ] &=& 1 - \frac{\tau}{L},
\eeq
as well as
\beq
\, [ N_{\tau} , \bar{b}_{k_0}] &=& -  \bar{b}_{k_0}, \quad  [ N_{\tau} , \bar{b}_{k_0}^{\dagger}] = \bar{b}^{\dagger}_{k_0}, \\
\, [ N_{\tau} , \tilde{b}_{k_0}] &=& [ N_{\tau} , \tilde{b}_{k_0}^{\dagger}] = 0.
\eeq
Accounting them in application of the Baker-Campbell-Hausdorff formula, we establish the identity
\beq
e^{i \chi N_{\tau}} e^{\kappa \alpha b_{k_0}^{\dagger}} = e^{\kappa \alpha \tilde{b}_{k_0}^{\dagger}} e^{\kappa \alpha z \bar{b}_{k_0}^{\dagger}} e^{i \chi N_{\tau}},
\label{id1}
\eeq
where $z=e^{i \chi}$. Next, we define $\bar{B} = \sqrt{\frac{L}{\tau}} \bar{b}_{k_0}$ and $\tilde{B} = \sqrt{\frac{L}{L-\tau}} \tilde{b}_{k_0}$, such that $[\bar{B}, \bar{B}^{\dagger}] = [\tilde{B}, \tilde{B}^{\dagger}]=1$,  and establish
\beq 
& & e^{\kappa \alpha^* b_{k_0}}  e^{\kappa \alpha \tilde{b}_{k_0}^{\dagger}} e^{\kappa \alpha z \bar{b}_{k_0}^{\dagger}} \nonumber \\
&=& e^{\kappa \alpha^*  \sqrt{\frac{\tau}{L}} \bar{B}} e^{\kappa \alpha z  \sqrt{\frac{\tau}{L}} \bar{B}^{\dagger}} e^{\kappa \alpha^* \sqrt{\frac{L-\tau}{L}}\tilde{B}} e^{\kappa \alpha \sqrt{\frac{L-\tau}{L}} \tilde{B}^{\dagger}}  \nonumber \\
&=&  e^{z \kappa^2 |\alpha|^2 \frac{\tau}{L} + \kappa^2 |\alpha|^2 \frac{L-\tau}{L}} e^{\kappa \alpha z  \sqrt{\frac{\tau}{L}} \bar{B}^{\dagger}} e^{\kappa \alpha^*  \sqrt{\frac{\tau}{L}} \bar{B}}  \nonumber \\
& & \times e^{\kappa \alpha  \sqrt{\frac{L-\tau}{L}} \tilde{B}^{\dagger}} e^{\kappa \alpha^* \sqrt{\frac{L-\tau}{L}}\tilde{B}} \nonumber \\
&=&  e^{z \kappa^2 |\alpha |^2 \frac{\tau}{L} + \kappa^2 |\alpha |^2 \frac{L-\tau}{L}} e^{\kappa \alpha z  \bar{b}^{\dagger}_{k_0}}   e^{\kappa \alpha \tilde{b}_{k_0}^{\dagger}} e^{\kappa \alpha^* b_{k_0}} .
\label{id2}
\eeq
Finally, using the identity analogous to \eq{id1}
\beq
e^{\kappa \alpha^* b_{k_0}} e^{i\chi N_{\tau}} = e^{i \chi N_{\tau}} e^{\kappa \alpha^* \tilde{b}_{k_0}} e^{\kappa \alpha^* z \bar{b}_{k_0}}, 
\label{id3}
\eeq
and combining the result together with \eq{id1} and \eq{id2}, we obtain the expression
\beq
& & e^{\kappa \alpha^* b_{k_0}} e^{i\chi N_{\tau}} e^{\kappa \alpha b_{k_0}^{\dagger}} = e^{z \kappa^2 |\alpha |^2 \frac{\tau}{L} + \kappa^2 |\alpha |^2 \frac{L-\tau}{L}} \nonumber \\
& & \times  e^{\kappa \alpha z  \bar{b}^{\dagger}_{k_0}}   e^{\kappa \alpha \tilde{b}_{k_0}^{\dagger}} e^{i \chi N_{\tau}} e^{\kappa \alpha^* \tilde{b}_{k_0}} e^{\kappa \alpha^* z \bar{b}_{k_0}},
\label{id4}
\eeq
whose rhs contains the exponents of annihilation (creation) operators standing to the right (left) from $e^{i \chi N_{\tau}}$, in contrast to the original reciprocal arrangement appearing in lhs of this expression. Inserting \eq{id4} into \eq{fcs_def}, we deduce
\beq 
& & F_{\tau}^{(\kappa)} (\chi)  e^{-(z-1) \kappa^2 |\alpha |^2 \frac{\tau}{L}}
\nonumber \\
&=& \langle 0 | S_0^{\dagger}  e^{z_{\kappa} \alpha  \bar{b}^{\dagger}_{k_0}}   e^{ \alpha \tilde{b}_{k_0}^{\dagger}} e^{i \chi N_{\tau}} e^{z_{\kappa} \alpha^*  \bar{b}_{k_0}} e^{ \alpha^* \tilde{b}_{k_0}} S_0 | 0 \rangle \nonumber \\
&=& \langle 0 | S_0^{a \, \dagger} [L/2,-L/2] 
 e^{\alpha b_{k_0}^{f \, \dagger}} e^{z_{\kappa} \alpha  \bar{b}^{\dagger}_{k_0}}   e^{\alpha b_{k_0}^{p \, \dagger}}  \nonumber \\
& \times & e^{i \chi N_{\tau}} e^{\alpha^* b_{k_0}^p} e^{z_{\kappa} \alpha^*  \bar{b}_{k_0}}  e^{\alpha^* b_{k_0}^f} S_0^a [L/2,-L/2]  |0 \rangle  \nonumber \\
&+&   \frac{|\lambda |^2}{2 \Gamma} \langle 0 | S_0^{b \, \dagger} [L/2,-L/2] 
 e^{\alpha b_{k_0}^{f \, \dagger}} e^{z_{\kappa} \alpha \bar{b}^{\dagger}_{k_0}}   e^{\alpha b_{k_0}^{p \, \dagger}}  \nonumber \\
& \times & e^{i \chi N_{\tau}} e^{\alpha^* b_{k_0}^p} e^{z_{\kappa} \alpha^*  \bar{b}_{k_0}}  e^{\alpha^* b_{k_0}^f} S_0^b [L/2,-L/2]  |0 \rangle , 
\label{fcs:fcp}
\eeq
where $z_{\kappa} = \kappa (z - 1) +1$. 

The result of counting should not depend on the ``future'' -- this is a manifestation of the causality principle. Therefore, we expect that \eq{fcs:fcp} identically equals
\beq
& &  F_{\tau}^{(\kappa)} (\chi) e^{-(z-1)\kappa^2 |\alpha |^2 \frac{\tau}{L}} \nonumber \\
&=&  \langle 0 | S_0^{a \, \dagger} [L/2,z_1]  e^{z_{\kappa} \alpha  \bar{b}^{\dagger}_{k_0}}   e^{\alpha b_{k_0}^{p \, \dagger}}  \nonumber \\
&\times & e^{i \chi N_{\tau}} e^{\alpha^* b_{k_0}^p} e^{z_{\kappa} \alpha^*  \bar{b}_{k_0}}   S_0^{a} [L/2,z_1]  |0 \rangle \nonumber \\
&+& \frac{|\lambda|^2}{2 \Gamma}  \langle 0 | S_0^{b \, \dagger} [L/2,z_1]  e^{z_{\kappa} \alpha  \bar{b}^{\dagger}_{k_0}}   e^{\alpha b_{k_0}^{p \, \dagger}}  \nonumber \\
& \times & e^{i \chi N_{\tau}} e^{\alpha^* b_{k_0}^p} e^{z_{\kappa} \alpha^*  \bar{b}_{k_0}}   S_0^{b} [L/2,z_1]  |0 \rangle .
\label{fcs:cp}
\eeq
This statement can be rigorously proven using the splitting relations \eq{split_alg}. The identities \eq{norm_id1}-\eq{norm_id3} appear essential for such a proof.

In the next step we integrate out the fields in the ``past''. They do influence the counting, therefore the result of this procedure can not be simply guessed. To eliminate the ``past'' fields in a systematic way, we employ the first and the second relations of \eq{split_alg}, splitting the interval $[z_1 ,L/2]$ into the counting $\tau = z_2 -z_1$ and the ``past'' (or waiting) $T=L/2-z_2$ subintervals. Applying the Wick's theorem separately on each subinterval, we obtain the formula \eq{fcs:c}.

\section{Laplace transforms \eq{LT}}
\label{LT_app}

The explicit expressions for the Laplace transforms \eq{LT}  are obtained by the resummation of geometric series in the Laplace space which emerge from the unique possibility for the Wick's contraction of the path-ordered field operators. The result of this procedure reads
\beq
\mathcal{G}_{aa} (p) &=& \mathcal{G}_{\bar{a} \bar{a}}  (p) = \frac{\tilde{r} (p; u, v^*)}{1 - |\lambda|^2 r (p; u, v^*)}, \\
\mathcal{G}_{bb} (p) &=& \frac{r (p; u, v^*)}{1 -|\lambda |^2 r (p; u , v^*)}, \\
\mathcal{G}_{cc} (p) &=& \tilde{\tilde{r}} (p; u , v^*) + \frac{|\lambda|^2 \tilde{r}^2 (p; u , v^*)}{1-|\lambda|^2 r (p; u , v^*)}, 
\eeq
\beq
\mathcal{G}_{\bar{a}b} (p) &=& \mathcal{G}_{a b} (p) = \frac{c (p; u, v^*)}{1 -|\lambda |^2 r (p; u , v^*)}, \\
\mathcal{G}_{b \bar{a}} (p) &=&  \mathcal{G}_{b a} (p)= \frac{\bar{c} (p; u, v^*)}{1 -|\lambda |^2 r (p; u , v^*)} , \\
\mathcal{G}_{ca} (p) &=&  \mathcal{G}_{c\bar{a}} (p)= b (p; u , v^*) \nonumber \\
& & \qquad + \frac{|\lambda|^2 c (p; u, v^*) \tilde{r} (p; u , v^*)}{1-|\lambda|^2 r (p; u , v^*)}, \\
\mathcal{G}_{ac} (p) &=& \mathcal{G}_{\bar{a}c} (p)= \bar{b} (p; u , v^*) \nonumber \\
& & \qquad + \frac{|\lambda|^2 \bar{c} (p; u , v^*) \tilde{r} (p; u, v^*) }{1-|\lambda|^2 r (p; u , v^*)}, 
\eeq
\beq
\mathcal{G}_{a\bar{a}} (p) &=& \tilde{r} (p; u , v^*) + \frac{|\lambda|^2 c (p; u, v^*) \bar{c} (p; u , v^*)}{1-|\lambda|^2 r (p; u , v^*)},
\eeq
\beq
\mathcal{G}_{c b} (p) &=& f (p; u , v^*) + \frac{|\lambda|^2 c (p; u, v^*) c (p; u , v^*)}{1-|\lambda|^2 r (p; u , v^*)}, \\
\mathcal{G}_{b c} (p) &=& \bar{f} (p; u , v^*) + \frac{|\lambda|^2 \bar{c} (p; u, v^*) \bar{c} (p; u , v^*)}{1-|\lambda|^2 r (p; u , v^*)},
\eeq
where the functions
\beq
r (p; u, v^*) &=&  \int_0^{\infty} d x \, d_u^* (x) d_v (x) e^{ip x},  \label{ruv} \\
\tilde{r} (p; u, v^*) &=&  \int_0^{\infty} d x \, \tilde{d}_u^* (x) \tilde{d}_v (x) e^{ip x}, \label{truv} \\
\tilde{\tilde{r}} (p; u, v^*) &=&  \int_0^{\infty} d x \, \tilde{\tilde{d}}_u^* (x) \tilde{\tilde{d}}_v (x) e^{ip x}, \label{ttruv} \\
c (p; u, v^*) &=&  \int_0^{\infty} d x \, \tilde{d}_u^* (x) d_v (x) e^{ip x},\label{cuv}  \\
\bar{c} (p; u, v^*) &=&  \int_0^{\infty} d x \, d_u^* (x) \tilde{d}_v (x) e^{ip x}, \\
b (p;u,v^*) &=& \int_0^{\infty} d x \, \tilde{\tilde{d}}_u^* (x) \tilde{d}_v (x) e^{ip x} , \label{buv} \\
\bar{b} (p;u,v^*) &=& \int_0^{\infty} d x \, \tilde{d}_u^* (x) \tilde{\tilde{d}}_v (x) e^{ip x}, \\
f (p; u, v^*) &=& \int_0^{\infty} d x \, \tilde{\tilde{d}}^*_u (x) d_v (x) e^{i p x} , \label{fuv} \\
\bar{f} (p; u, v^*) &=& \int_0^{\infty} d x \, d^*_u (x)  \tilde{\tilde{d}}_v (x) e^{i p x}
\eeq
are obtained using the explicit form of the functions $d_v (x)$, $\tilde{d}_v (x)$, $\tilde{\tilde{d}}_v (x)$ from Eqs.~\eq{dv},\eq{tdv},\eq{ttdv}.

\section{Computation of \eq{lam_aa_res}-\eq{lam_ca_res}}
\label{app:norm}

To compute the quantities in \eq{lam_aa_res}-\eq{lam_ca_res}, we insert into \eq{lam_fin} the expressions for $\mathcal{G}_{\beta' \beta} (p)$ from the Appendix \ref{LT_app} and obtain
\beq
& & \Lambda_{aa} (\tau) + \frac{|\lambda|^2}{2 \Gamma} \Lambda_{bb} (\tau) \label{lam_aa_def} \\
&=& \int \frac{d p}{2 \pi}  e^{-i p \tau} \frac{\tilde{r} (p; z_{\kappa} \alpha, z_{\kappa} \alpha^*) +\frac{|\lambda |^2}{2 \Gamma} r (p; z_{\kappa} \alpha, z_{\kappa} \alpha^*) }{1 -z |\lambda|^2 r (p; z_{\kappa} \alpha, z_{\kappa} \alpha^*)} , \nonumber 
\eeq
\beq
& & \Lambda_{cc} (\tau)  + \frac{|\lambda|^2}{2 \Gamma}  \Lambda_{\bar{a} \bar{a}} (\tau)  \label{lam_cc_def} \\
&=& \int \frac{d p}{2 \pi}  e^{-i p \tau} \left[ \tilde{\tilde{r}} (p; z_{\kappa} \alpha, z_{\kappa} \alpha^*)+ \frac{|\lambda|^2}{2 \Gamma} \tilde{r} (p; z_{\kappa} \alpha, z_{\kappa} \alpha^*) \right. \nonumber \\
& & \left. \qquad \qquad \times \frac{1 + 2 \Gamma z \tilde{r} (p; z_{\kappa} \alpha, z_{\kappa} \alpha^*)}{1 - z |\lambda|^2 r (p; z_{\kappa} \alpha, z_{\kappa} \alpha^*)} \right],  \nonumber
\eeq
\beq
& & \Lambda_{ca} (\tau)  + \frac{|\lambda|^2}{2 \Gamma}  \Lambda_{\bar{a} b} (\tau)  \label{lam_ca_def} \\
&=& \int \frac{d p}{2 \pi}  e^{-i p \tau} \left[ b (p; z_{\kappa} \alpha, z_{\kappa} \alpha^*)+ \frac{|\lambda|^2}{2 \Gamma} c (p; z_{\kappa} \alpha, z_{\kappa} \alpha^*)\right.  \nonumber \\ 
& & \left. \qquad \qquad \times \frac{1 + 2 \Gamma z \tilde{r} (p; z_{\kappa} \alpha, z_{\kappa} \alpha^*)}{1 - z |\lambda|^2 r (p; z_{\kappa} \alpha, z_{\kappa} \alpha^*)} \right].
\nonumber
\eeq

Establishing the functions \eq{ruv}, \eq{truv}, \eq{ttruv}, \eq{cuv}, \eq{buv} for $u=z_{\kappa} \alpha$ and $v^* = z_{\kappa} \alpha^*$
in the explicit form
\beq
& &  r (p; z_{\kappa} \alpha, z_{\kappa} \alpha^*) = - \frac{2 i (\delta^2 +\Gamma^2) (p+i \Gamma)}{Q_0 (p, z_{\kappa})} , \label{r_expl} \\
& & \tilde{r} (p; z_{\kappa} \alpha, z_{\kappa} \alpha^*) = i \frac{R_0 (p)  -   \frac{\Omega_r^2}{2} (z_{\kappa}   - 2) (p+i \Gamma)}{Q_0 (p, z_{\kappa})} , \label{rt_expl} 
\eeq
\beq
& & \tilde{\tilde{r}} (p; z_{\kappa} \alpha, z_{\kappa} \alpha^*) \label{rtt_expl}  \\
& & \qquad = \frac{i}{Q_0 (p, z_{\kappa})}  \left[ R_0 (p) -  i z_{\kappa} \Gamma \Omega_r^2  \frac{ p^2 +\delta^2 +\Gamma^2}{2 (\delta^2 +\Gamma^2)} \right. \nonumber \\
& & \left.  + \Omega^2_r (p+i \Gamma) \left(1  - z_{\kappa} \frac{\delta^2 -\Gamma^2}{\delta^2 +\Gamma^2}  - z_{\kappa}^2 \frac{ \Omega_r^2 }{ 8 (\delta^2 +\Gamma^2) } \right)  \right] ,
\nonumber 
\eeq
\beq
& & c (p; z_{\kappa} \alpha, z_{\kappa} \alpha^*) \label{c_expl} \\
& &  \qquad =\frac{i(\delta +i \Gamma) (p+2 i \Gamma) (p-\delta +i \Gamma) }{Q_0 (p, z_{\kappa})} , \nonumber\\
& & b (p; z_{\kappa} \alpha, z_{\kappa} \alpha^*) = \tilde{r} (p; z_{\kappa} \alpha, z_{\kappa} \alpha^*)  \label{b_expl}\\
& & \qquad -\frac{i \Omega_r^2 z_{\kappa} (p+2 i \Gamma) (p+\delta +i \Gamma)}{4 (\delta - i \Gamma) Q_0 (p, z_{\kappa})} , \nonumber
\eeq
where 
\beq
Q_0 (p, z_{\kappa}) &=& p R_0 (p) - i \Gamma \Omega_r^2  ( p + i \Gamma)  \nonumber \\
& & - (z_{\kappa}-1) \Omega_r^2 (p + i \Gamma)^2 ,
\eeq
and inserting them into \eq{lam_aa_def}-\eq{lam_ca_def}, we arrive after some transformations at \eq{lam_aa_res}-\eq{lam_ca_res}.

\section{Correlation functions \eq{G_def0}}
\label{app:rcmn}

Setting $u=v= \alpha$, we introduce the following notations
\beq
\bar{\mathcal{G}}_{bb} (\tau) &=&   R (\tau) , \\
\bar{\mathcal{G}}_{ab} (\tau) &=&   \bar{\mathcal{G}}_{ba}^* (\tau) = C (\tau) , \\
\bar{\mathcal{G}}_{a \bar{a}} (\tau) &=& M (\tau), \\
\bar{\mathcal{G}}_{cb} (\tau) &=&  \bar{\mathcal{G}}_{bc}^* (\tau)= N (\tau) .
\eeq
From the normalization identities \eq{norm_id1}-\eq{norm_id3} it follows
\beq
\bar{\mathcal{G}}_{aa} (\tau) &=&  \bar{\mathcal{G}}_{\bar{a}\bar{a}} (\tau)= 1  - \frac{|\lambda |^2 }{2 \Gamma} R (\tau), \\ 
\bar{\mathcal{G}}_{cc}  (\tau) &=&  1 + \frac{|\lambda |^4}{4 \Gamma^2}  R (\tau) , \\
\bar{\mathcal{G}}_{ca} (\tau) &=&  \bar{\mathcal{G}}_{ac}^* (\tau)=  1 - \frac{|\lambda|^2}{2 \Gamma} C (\tau) .
\eeq

We establish the following differential equations for the functions $R (\tau)$, $C (\tau)$, $M (\tau)$, and $N (\tau)$.

The function $R (\tau)$ obeys the differential equation
\beq
\dddot{R} (\tau) + 4 \Gamma \ddot{R} (\tau) + \left( \Omega_r^2+ \delta^2 +5 \Gamma^2\right) \dot{R} (\tau) \nonumber\\
+ \Gamma \left( \Omega_r^2 + 2 \delta^2 +2 \Gamma^2\right) R (\tau) = 2 \Gamma \left( \delta^2 +\Gamma^2 \right), \label{diff3}
\eeq
equipped with the initial conditions $R (0) = \dot{R} (0) = 0$, $\ddot{R} (0) = 2 (\delta^2 +\Gamma^2)$.

The function $M (\tau)$ obeys the same differential equation \eq{diff3} equipped the initial conditions $M (0)=1$, $\dot{M} (0) = 0$, 
$\ddot{M} (0) = -  \Omega_r^2 /2$.

The function $C (\tau)$ obeys the differential equation
\beq
& & \dot{C} (\tau) -i (\delta + i \Gamma) C (\tau) \nonumber \\
&=& -i (\delta + i \Gamma) \left( 1 - \frac{\Omega_r^2}{2 ( \delta^2+\Gamma^2)} R (\tau) \right),
\eeq
with the initial condition
$C (0) =0$.

The function $N (\tau)$ obeys the differential equation
\beq
& & \dot{N} (\tau) - i (\delta + i \Gamma ) N (\tau) \nonumber \\
&=& - \frac{\delta + i \Gamma}{\delta - i \Gamma}  \left(     \dot{M} (\tau)   + i (\delta - i \Gamma) M (\tau) \right), 
\eeq
with the initial condition $N (0) =0$.

At small $\tau$ these functions have the following behavior
\beq
R (\tau) & \approx & (\delta^2 +\Gamma^2 ) \tau^2 ,\nonumber \\
C ( \tau ) & \approx & - i (\delta + i \Gamma) \tau ,\nonumber \\
M (\tau) & \approx & 1 - \frac{\Omega_r^2 \tau^2}{4} ,\nonumber \\
N (\tau) & \approx & - i (\delta +i \Gamma) \tau , \nonumber
\eeq
while in the limit $\tau \to \infty$ they all reach the same stationary value $\frac{\Gamma}{\Gamma + |\lambda|^2} = \frac{2 \delta^2 + 2 \Gamma^2 }{2 \delta^2 + 2 \Gamma^2 + \Omega_r^2}$.

\begin{figure}[t]
    \includegraphics[width=0.48\textwidth]{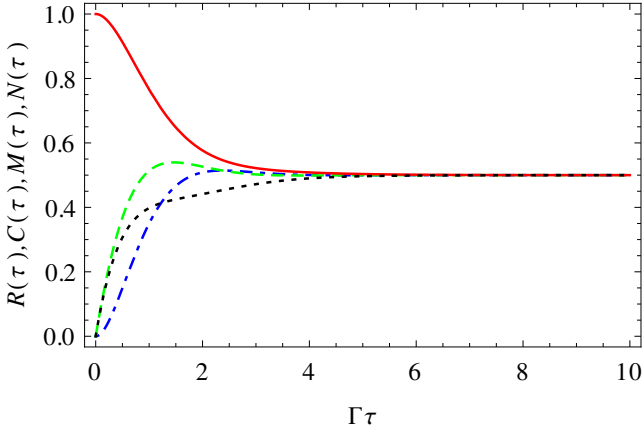}
    \caption{(Color online) The functions $R (\tau)$ (dot-dashed blue), $C (\tau)$ (dashed green), $M (\tau)$ (solid red), and $N (\tau)$ (dotted black), for $\delta = 0$ and $\Omega_r = \sqrt{2}\Gamma$. At large $\tau$ they all saturate at the value 0.5.} \label{fig:rcmn}
\end{figure}

In the resonant case $\delta=0$ we find the analytic solutions  to these differential equations
\beq
R (\tau) &=& \frac{2 \Gamma^2}{\Omega_r^2 + 2 \Gamma^2} \\
&-& \frac{2 \Gamma^2 (\bar{\Omega}_r \cos \bar{\Omega}_r \tau+ \frac{3 \Gamma}{2} \sin \bar{\Omega}_r \tau  ) }{\bar{\Omega}_r (\Omega_r^2 +2  \Gamma^2 )}  e^{- \frac{3 \Gamma}{2} \tau} , \nonumber\\
C (\tau) &=& \frac{2 \Gamma^2}{\Omega_r^2 + 2 \Gamma^2} \\
&+&  \frac{\Gamma [ (\Omega^2_r - \Gamma^2) \sin \bar{\Omega}_r \tau - 2 \Gamma \bar{\Omega}_r \cos \bar{\Omega}_r \tau]}{ \bar{\Omega}_r (\Omega_r^2 + 2 \Gamma^2 )}  e^{- \frac{3 \Gamma}{2} \tau}, \nonumber
\eeq
\beq
M (\tau) &=&  \frac{2 \Gamma^2}{\Omega_r^2 + 2 \Gamma^2} + \frac12 e^{-\Gamma \tau}  \\
&+& \frac{\Omega_r^2 -2 \Gamma^2}{2 (\Omega_r^2 +2 \Gamma^2)} e^{- \frac{3 \Gamma}{2} \tau} \cos \bar{\Omega}_r \tau \nonumber\\
&+& \frac{\Gamma}{4 \bar{\Omega}_r} \frac{5 \Omega_r^2 - 2 \Gamma^2}{\Omega_r^2 +2 \Gamma^2} e^{- \frac{3 \Gamma}{2} \tau} \sin \bar{\Omega}_r \tau , \nonumber  \\
N (\tau) &=& M (\tau) - e^{-\Gamma \tau},
\eeq
where $\bar{\Omega}_r = \sqrt{\Omega_r^2 - \frac{\Gamma^2}{4}}$. They are shown in Fig.~\ref{fig:rcmn} for $\Omega_r = \sqrt{2}\Gamma$.

\end{document}